\documentclass[aps,prl,twocolumn,superscriptaddress,groupedaddress]{revtex4}
\usepackage{subfig}
\usepackage{graphicx}  
\usepackage{dcolumn}   
\usepackage{bm}        
\usepackage{amssymb}   
\usepackage{slashed}
\usepackage{graphicx}				
\usepackage{amsmath}
\usepackage{mathtools}
\usepackage{tikz,pgf}
\usepackage{tikz-cd}
\usepackage{comment}
\usepackage{asymptote}
\usetikzlibrary{arrows,backgrounds}
\usetikzlibrary{fit,scopes,calc,matrix,positioning,decorations.pathmorphing}
\usepackage[all]{xy}
\usepackage{yfonts}
	
\usepackage{xcolor}

\newcommand{\Bracket}[1]{\ensuremath{\left\langle#1\right\rangle}}

\DeclareFontFamily{OMS}{oasy}{\skewchar\font48 }
\DeclareFontShape{OMS}{oasy}{m}{n}{%
         <-5.5> oasy5     <5.5-6.5> oasy6
      <6.5-7.5> oasy7     <7.5-8.5> oasy8
      <8.5-9.5> oasy9     <9.5->  oasy10
      }{}
\DeclareFontShape{OMS}{oasy}{b}{n}{%
       <-6> oabsy5
      <6-8> oabsy7
      <8->  oabsy10
      }{}
\DeclareSymbolFont{oasy}{OMS}{oasy}{m}{n}
\SetSymbolFont{oasy}{bold}{OMS}{oasy}{b}{n}

\DeclareMathSymbol{\smallleftarrow}     {\mathrel}{oasy}{"20}
\DeclareMathSymbol{\smallrightarrow}    {\mathrel}{oasy}{"21}
\DeclareMathSymbol{\smallleftrightarrow}{\mathrel}{oasy}{"24}

\begin{document}
\title{The hidden quantum origin of gauge connections }
\author{Andrei T. Patrascu}
\address{ELI-NP, Horia Hulubei National Institute for R\&D in Physics and Nuclear Engineering, 30 Reactorului St, Bucharest-Magurele, 077125, Romania\\
email: andrei.patrascu.11@alumni.ucl.ac.uk}

\begin{abstract}
A fibre bundle viewpoint of gauge field theories is reviewed with focus on a possible quantum interpretation. The fundamental quantum properties of non-separability of state spaces is considered in the context of defining the connection on the fibre bundle, leading to an application of the quantum principles to the geometrical and topological definition of gauge theories. As a result, one could justifiably ask oneself if all interactions of the standard model, and perhaps even classical gravity have some quantum component after all. I employ a standard fibre bundle approach to introduce gauge theories, albeit it is known that a quantum bundle exists, simply because the main scope is to show that in the usual way in which we formulate classical gauge theories one can find quantum aspects that have been unknown until now. In a sense, I will try to justify the assessment that if we are to allow for gauge fields and parallel transport, we may have to allow at least some level of quantumness even in our classical gauge theories. The main statement is that propagation of interactions in spacetime is a quantum phenomenon.
After writing the first draft of this article I noticed ref. [13] where the authors device entanglement of what they call "classical light". This experiment supports my theoretical developments with the distinction that I interpret such phenomena also as fundamentally quantum. The distinction comes from the fact that the quantum nature of the experiments is manifested in a different way. My view on this is that there is no purely classical reality, no matter what the scale is at which we consider the description. 
I also discuss the fact that observing a quantum nature of "classical" light propagation would amount to the requirement of modifying the causal structure defined in terms of the speed of light in a vacuum, on stronger grounds, based on the quantum interpretation of gauge connections.
\end{abstract}
\maketitle

\section{introduction}
The first quantum field theories were free theories. It was clear how to describe free quantum fields, travelling through space, but for the interaction to be possible, one had to implement the idea of gauge invariance and by extension, of gauge connection [1]. Gauge symmetries are therefore absolutely required for a theory to have meaningful interactions, so, despite the fact that they were regarded as "spurious" or "superfluous" it was hard to make sense of anything without them. But, while gauge symmetry offered us the idea of interaction, it also appeared to emerge from a geometric, fibre bundle definition of a connection, a method through which one connects adjacent regions on a manifold [2]. In fact, in order to create a G-bundle interpretation of gauge theory, we need, as for any G-bundle, to have a set of transition functions 
\begin{equation}
g_{ij}: U_{i}\cap U_{j} \rightarrow G
\end{equation}
associating to any overlap in the base space $M$ a group element in $G$. If we consider a triple intersection, then these transition functions satisfy the cocycle condition leading to a cohomological interpretation of the process. It is important to see that this process has an analogue in the foundations of quantum mechanics. Quantum mechanics in its most fundamental and simple form is constructed based on the assumption that the cartesian product of spaces of states is not sufficient to describe a combined system made out of systems each with state spaces constructed only from the two parts separately [3-6]. Take for example two such systems, $X$ and $Y$ and then their respective state spaces $E$ and $F$. The composite system "$X$ and $Y$" has a space of states which is not always the cartesian product of the individual spaces of states, $E\times F$. In fact, in quantum mechanics, which obeys the rules of a $Hilb$ category and involves linear transformations as maps, the product of the individual state spaces is $E\otimes F$. This space is larger and involves entangled states that do not belong to any subsystem while being characteristic to the composite system as a whole. This induces a tension between a global structure and a local structure, namely if the two (or more) subsystems are taken separately, one obtains a cartesian product state, but with no interaction between them, a free theory, while if one combines them one introduces interactions, but one also reaches into the global structure of the combined system which has properties that cannot be recovered locally.  Not surprisingly, to compensate for this tension between local and global (or neighbouring) patches we introduce connections which are in gauge theories, precisely the interactions.  These interactions would not exist if we had a strictly cartesian way of patching up the manifold. Therefore interaction comes with at least some minimal quantum effect (entanglement). The solution is for the fibre bundle approach to take the intersection between two patches and to establish transition functions which connect those patches with our gauge group. The freedom to make choices in this gauge group for transformations of frames in the intersection of the patches, so that we can translate our data from one patch to the next, is what allows us to construct theories with interactions. The same mechanism appears naturally in quantum mechanics in the phase of our quantum wavefunctions and fields. This is why it is only too natural to make changes in those phases which are impossible to detect directly (but have the well known effects of quantum interference, etc. hence the name "undetectable" may be a bit of an exaggeration). Those changes are indeed impossible to detect locally, or in any single measurement of a quantum system, but become extremely visible once we go beyond the single measurement and take a statistics that covers more regions of our manifold [7]. This is the path integral quantisation approach, that gives us access to the global structure of our space. 
The transition functions connect our base manifold patches, particularly their intersections, to the gauge group $G$ and encode some additional information which may ultimately translate into the global information that describes our bundle. That global information binds the two patches, and due to the global structure (curvature) it is not reducible to any of the patches anymore. Whenever we try to project on the independent patches, we lose the global information that facilitates the propagation of interaction. Of course, global information does not need to be bound to spacetime alone (although, for gravitation the space with curvature will be spacetime itself and it will be it that will carry the global information). In the case of the other interactions spacetime may be flat, but we still have global structure, in the form of curvature, in the gauge space; after all, this is what the field strength $F_{\mu\nu}$ is. Propagation of interaction therefore involves some global structure that contains information that is non-reducible to local patches, hence, it is fundamentally quantum. 
Therefore there are at least two different ways in which the global structure can be probed. First, it would be through the gauge structure of our interactions. For example, in gravity the observables are non-local, as they are described starting with a curvature form which naturally must include some global structure. In non-gravitational interactions the non-locality manifests itself in the gauge field space, through the same type of curvature, the only difference being its realisation: in gravity the non-local observables are space-time based, in the other gauge interactions they are gauge-space based. 
The other method allowing us to access the global structure of our manifold appears in quantum mechanics as the complex phase structure of the probability amplitude. While in classical physics the probability of a series of events leading from $A$ to $C$ having the outcome $a$ for $A$ and $c$ for $C$ going through an intermediate state $B$ having the outcome $b$ is calculated as $P_{abc}=P_{ab}P_{bc}$ and given several intermediate outcomes, we calculate $P_{ac}=\sum_{b}P_{abc}$, in quantum mechanics we cannot assume that there is an intermediate measurement that fixes the outcome $b$ for $B$. Therefore we introduce a complex construction $P_{ac}^{q}=|\phi_{ac}|^{2}$ with $\phi_{ac}=\sum_{b}\phi_{ab}\phi_{bc}$ which, by means of its imaginary part correlates paths from outside the classical correlation region. The fact that there is no pre-defined outcome for an intermediate observable allows for broader correlations reaching into the global structure of our manifold. 
It is interesting to note that these two approaches should finally be providing equivalent descriptions of the global structure, hence they may be mathematically linked. I will speculate that the type of link is similar to the connection between the topological index and the analytic index in the Atiyah-Singer index theorem and its generalisations. At the same time this approach may open new links between harmonic analysis (Fourier transforms as a building block for entanglement [12]), combinatorics (the structure of the partition function and the Fuglede conjecture) and geometry (curvature).

It is interesting to note that there exists always a relation between local and global structures. While this relation is often not one of perfect determination, there are various properties that a global structure uniquely determines upon a local structure. This approach has been taken in the Morse theory to find out new connections between the differential functions on a manifold and the global topological structure of that manifold. 
It is also worth noting that in physics in general, there are certain limitations in the connection between global and local structures. For example, due to the causal structure of spacetime, or the locality of interactions, it is impossible to have a direct signal communicating global (or otherwise stated, remote, space-like separated) properties of the manifold to the local observer. Local Lorentz symmetry therefore gives us a barrier from receiving a-causal information. However, geometry and topology seem to relate local and global properties in various ways that are mathematically well established. Quantum mechanics indeed has a series of fundamental uncertainties that cannot be eliminated, but those uncertainties allow it (and are a consequence of the ability) to probe the global structure of the manifold it is constructed in, and therefore, by analysing statistical results, a local observer may infer the existence of non-trivial global structures. If quantum mechanics relied only on classical probability, with no complex phase or probability amplitudes, such a connection between local and global would not have been possible, thinking only in terms of signalling and spacetime causality. While it is clear that quantum non-locality does not imply a-causal connections or non-local interactions, it is a way in which the effects of the global structure become detectable locally. 
As an analogy therefore, in the same way in which global information requires the construction of curvature terms, in order to take into account the non-triviality of connections over our spacetime based fibre bundle, in quantum mechanics global information is recovered by the fact that multiple paths have to be considered in the path integral formulation in order to find the proper statistical behaviour. I suspect this analogy can be made more precise, but I leave this for a future article.
For now let us look back at our fibre bundle. The transition functions, having the structure of a cocycle and having global properties amenable to a homological analysis, give us the link between local and global structure. Our patches $\{U_{i}\}$ considered as trivialising neighbourhoods if they cover M, are local trivialisations, and link our base manifold with the fibres by means of a cartesian product
\begin{equation}
\phi_{i}:\pi^{-1}(U_{i}) \rightarrow U_{i} \times F
\end{equation}
where $x\in M$ and $\pi^{-1}(x)$ is a fibre over $x\in M$. 
If we have a fibre $\pi^{-1}(x)$ given, we can safely ignore the patch component and work exclusively with the fibre $F$ as in $f_{i}:\pi^{-1}(x) \rightarrow F$. This property called local trivialisation shows that our bundle is locally a cartesian product space. However, in general it is not a globally trivial space. This is why fibre bundles are used to describe topological i.e. non-local features. Quantum mechanics is entirely based on non-local features, albeit one has to underline that it is fundamentally a local theory in any causal relativistic sense[8]. The non-locality in quantum mechanics is probably a misnomer, but it is a way of thinking that the integration over all possible states/paths done via the path integral approach gives us access, albeit only in terms of statistical interferences, to the global structure of the manifold we are analysing. This also gives rise to non-trivial correlations, which we call entanglement, simply because they rely on the impossibility to reduce the global structure trivially back to a local decomposed one [9]. 

Our gauge theory, described in terms of fibre bundles, in fact has a global structure, defined by its curvature. The curvature tensor is the field strength, for example, of our electromagnetic theory. That does not mean that the curvature we have in gauge theories is the same as the one in general relativity, the differences are obvious when one looks at how the two quantities emerge in the respective equations of motion, one in degree one, the other in degree two. However, the main goal of this article is not to unify the two theories in any naive way that would emerge from such an analogy. What I consider to be interesting is that quantum properties can be recovered in the general procedure of generating gauge connections and hence, an axiomatisation of quantum mechanics would provide a better insight into how gauge interactions emerge. In fact, it seems likely that interactions by themselves require the type of local to global connection that only quantum mechanics can provide, namely a way in which subsystems can only be brought together via tensorial instead of cartesian products, and hence one cannot ignore the fact that the combination of two adjacent patches brings our system outside the cartesian product of the two patches taken separately. In that sense, it seems that even general relativity is to a very small extent, and in a hidden way, "quantum". Again, this is not an attempt at unifying quantum and general relativity, as a theory of quantum gravity would require most likely extended objects (strings). However, one should not overlook the idea that some quantum "remnants" could have slipped into the construction of general relativity from the way it has been constructed via connections on curved manifolds. It is therefore extremely important to understand the axiomatic structure of quantum mechanics [9]. For example, one may ask, what is fundamentally quantum? The expansion to observables that take the form of non-commuting matrices? The linear combination of vectors and the Born rule used to describe observations as emerging from the wavefunction? The fact that there is no pre-defined absolute outcome of an intermediate state in a process? The $Hilb$ category structure combining the Hilbert space with the operation of tensor product? All of those results are intrinsic to quantum mechanics of course, but they are not all independent. The inability of finding absolute outcomes for unmeasured intermediate states originates from the non-commutativity of the observables and their promotion to matrices. Entanglement and non-reducibility can also be linked to the idea of wavefunctions which is also required by the promotion of observables to matrix valued operators. While we know what is important in quantum mechanics, we are not fully certain yet about what axiomatises the theory. If the replacement of the cartesian product with the tensor product is indeed one aspect that defines quantum mechanics, as I consider it is, then that would sublimate to the idea that a global system created from the combination of subsystems generates a wider space of states than the cartesian combination between the two. 
All gauge theories have this component, as all have local gauge symmetries that define the connection and hence the gauge fields. In a curved space, be it spacetime or the inner space of a gauge field theory the situation is the same: we have linear transformations and linear combinations that transform from one patch to the next. In a sense, one can see this by means of the similarity of the Einstein index notation and the linear bra-ket notation commonly used in quantum mechanics. This linear structure and the associated tensor product structure shows that a common element is the fact that to transition from one patch to the next, one cannot rely simply on patching the two regions together by cartesian products, even if we assume non-quantum theories on both patches. The result is, as mentioned above, that when we construct classical theories as gauge theories, we introduce one aspect of quantum mechanics, which leads to an apparently strange behaviour of our classical theories, particularly when one tries to reconcile local and global properties. 
From a mathematical point of view, the mechanisms behind the construction of G-bundles appear to have several properties physicists would call "quantum", albeit those have never been described as such. 
It is also particularly interesting to note that if the two effects come in one package, namely the connection on fibres that links local information between adjacent patches and at the same time reaches into the global structure of the manifold, the idea of the classical causal cone that constructs our well known causal structure on which special relativity is based also has a quantum mechanical origin, being somehow the nearest neighbour "condensation" of a broader, more general structure suitable for the non-classical correlation between quantum fluctuations appearing somewhat outside the reach of a construction based solely on connecting adjacent patches. Indeed, that is what the cohomological structure of the transition functions allows us to access. This also leads us to quantum gravitational non-localities so often encountered in string theory. The causal connection seen in all gauge theories is a result of the propagation of local information, essentially at the speed of light in the vacuum, except for the case of massive carriers, which would then account for weak interactions and would bring us inside the causal cone. This whole structure however appears only as a first neighbour approximation of the mechanism generating the gauge connection. The same connection function also contains broader global information about the manifold, allowing us to reach into topological structures otherwise invisible. 

\section{gauge is quantum}
To make the statements of the introduction more precise, it is important to underline some general principles of mechanics and to discuss two apparently disconnected subjects: the gauge freedom and the quantum phase. It appears that quite generally, the universe gives us access to more variables than those needed for the description of the dynamics in all physical cases that are of practical importance. I always considered that somewhat strange, but the general assumption is that the additional degrees of freedom are always spurious and could, in principle, be eliminated. It is interesting to notice however that nobody did that and all advances, including the description of interactions and quantum mechanics appeared by expanding the number of variables and not decreasing them. 
To better understand the current situation let us have a look at a distinction we make in modern physics between observables and non-observables. In general, we consider an observable to be the equivalent of a question we may ask upon an experimental setup. We wish for example to find out the momentum of a massive particle, so we design an experiment where this observable makes unambiguous sense, allow the particle to get to a certain momentum, and then measure it. The observable is the momentum in the context of the experiment we designed. We have to be careful to notice that this observable cannot in general be defined outside a specific context in which it can be measured and determined even if only partially. The mathematical description for this is a hermitian, gauge invariant operator represented usually as a matrix in a certain basis. This means that given the experimental context the momentum we determine is a real physical number, and that various observers will agree upon the value measured. The agreement upon the measured value is usually encoded via the demand of gauge invariance. This agreement between observers played an important role in physics precisely because general solutions of mechanical systems do not in general describe only the system in some form of pure, context-free representation. Quite the opposite, solutions to mechanical equations come in most cases together with arbitrary functions which we need to add along the pure, context free solutions. The values of these functions at various times can indeed be arbitrary, and they do depend on arbitrary choices made by any observer. It does sound strange to say that solutions of dynamical problems in mechanics depend on arbitrary choices of observers, until we note that such choices may be just the choice of a reference frame. We know from the basics of Galilean mechanics that observable states of motion should not depend on choices of reference frames, a principle that has admirably been generalised in special relativity and then in general relativity. This independence of, for example, the speed of light, on the reference frames used, resulted in the postulate of relativity claiming that indeed the speed of light in a vacuum is independent on the choice of the speed of a reference frame. This resulted in the constancy of the speed of light in vacuum. As further developments in the theory of interactions emerged, it became clear that other invariances must be considered in order to be able to describe physical observable results unambiguously. We noticed that aside the arbitrary functions propagated along the solutions that allowed us to choose reference frames, we have a series of other arbitrary constructs, for example diffeomorphism invariance in general relativity, and gauge choice invariance in general interacting theories. 
However, the importance of gauge invariant observables has been somewhat exaggerated at the expense of the gauge-variant observables. After all, a gauge variant observable also represents a question that can be asked to an experimental setup, about a property of it or of its parts, and the outcomes can also be strictly speaking measured, the only difference is that observers that made different arbitrary choices of the gauge will observe different results. That doesn't make those results any less significant. To make this more precise let me go to two of the best known examples of this: general relativity and quantum chromodynamics. 
We know that all observables in general relativity that are to be gauge invariant must be non-local. That means in general relativity there is no unambiguous description of position. This makes perfect sense given that curvature can behave dynamically and can alter the local definition of position in a global (or non-local) way. To make an observable in general relativity gauge invariant (hence unambiguous to arbitrary choices of the observers) we have to dress it with non-local contributions, for example from Wilson loops. Therefore observables in general relativity are not localisable. However, in the limit case in which curvature is small, we can detect and approximately agree upon localised position, hence the observables can become approximately local and approximately gauge invariant, although never exactly so. 
Another situation happens to a theory that can in principle be fully spacetime local, but where the same problem arises, with the difference that observables cannot be both gauge invariant and local in another space, its inner gauge space. That theory is quantum chromodynamics. Indeed, we know that the colour of a particle is not a gauge invariant property. A different choice of gauge may change it and therefore observers outside would not be able to agree upon, say, what colour a specific quark has. This is why all free states we can observe are colour neutral, either in the form of hadrons made up of a neutral combination of three quarks, or in terms of various mesons made up of neutral combinations of colour and anti-colour. There again, for an observable to be purely gauge invariant, that means, unambiguously defined and robust to changes in arbitrary choices of functions of gauge, it must be non-local in the inner space where it has been defined. 
Up to now, the only use of gauge freedom was to make context choices related strictly to arbitrary choices of frames. We sort of expected that such choices of frames would not be important as we got used with this idea from both Galilei and Einstein. However, quantum mechanics makes the choice of arbitrary functions of the context even more stringent. In fact I will show that the gauge freedom expressed as arbitrary functions in solutions of dynamical problems emerges from the same considerations as the phase part of the wavefunction in quantum mechanics. In quantum mechanics the problem of observability is far more interesting because it involves other properties of a system and the arbitrary functions bring up not only correlations that are stronger than any classical correlation, but also refer to properties of the system that we expected to be well determined given our immediate experience. 
How an arbitrary choice of the value of a function that must be included in the solution affects the correlation in a system can be seen in the following simple imaginary experiment: 
let us consider the decay of a particle of spin zero into two particles of spin 1/2. Given the usual conservation laws, the two resulting particles have opposite spins. Therefore when we measure the spin of one, we will know instantly the spin of the other, no matter how far that other particle will be from the first. But the statement above is incomplete and ultimately wrong in at least two ways: first, to make a statement about spins being opposite we need to measure a spin projection, the spin remaining the same, 1/2. Second, to make a statement about a spin projection we need an axis, which we never provided in the setup above. The axis is the arbitrary function now, but its choice must have a clear effect on the system: it determines the state of the spin projection of being either $+1/2$ or $-1/2$. Once an axis is chosen, say, the $z$ axis, we can measure the spin projection of one electron, and obtain say $-1/2$ and we will be certain the other one electron measured on a parallel axis along $z$ will be $+1/2$. However, while performing measurements for our statistical evaluation, we can decide to turn the axis around, making now $x$ our projection axis. We measure the first electron on this new axis and obtain, say $+1/2$. On a parallel axis $x$ at the position of the other electron we will know for sure that the projection will be $-1/2$. This makes for the outcome of the experiment with our second choice of an arbitrary function for the axis. As we can see, the fact that we have an arbitrary function in our system propagating along the solutions, allows for a type of correlation not only between realised spin projection states, but between all non-realised and not determined projection states even before we introduced a choice for the axis. In this way the correlation is much higher than any classical correlation would allow (if by classical we mean only correlation between factually realised states). The constraint for the "spins" to be opposite, no matter what the spin projection axis would be, is a global, non-separable constraint on the system and in quantum mechanics we call that entanglement. The origin of gauge freedom and gauge arbitrary functions however is the same as the origin of the arbitrariness of choosing an axis in quantum mechanics, or the origin of a quantum complex phase. Expanding the concept of entanglement to gauge theories and their arbitrary functions results in constraints on the types of gauge connections and ultimately to the causal structure we know in special relativity. 
There is another discussion that deserves some consideration, namely the choice of gauge variant observables and the interpretation of some properties of the system. In general, we can form non-ambiguous observables in different ways depending on the context in which we ask our questions. At the same time, some non-ambiguous (gauge invariant) observables in one context may become totally gauge variant (ambiguous) in another context. 
A most amusing argument was brought by the idea that classical (i.e. large enough, in some interpretations, although I do not agree with that assessment) objects may not exist unless they are "observed". This would bring into the focus the "act of measurement" as important by itself. Even some physicists have argued that "the moon does not exist unless it's observed" and pointed out that this statement seems absurd. In fact it really is, but we have to distinguish the existence of an experimental setup and its context (the moon orbiting earth) from the existence of various properties that we expect such a setup to have, but that it doesn't actually have. The moon orbiting earth will continue to exist in the same way an electron in an atom continues to exist. About its properties, we can debate however. In this sense, the moon orbiting earth will have a series of well defined properties, like angular momentum around the earth, a mass, etc. All those observables will be gauge invariant and well defined quantum mechanically. But that doesn't mean quantum mechanics doesn't affect the orbit of the moon. The scale of the problem doesn't make it any less quantum, in the same sense in which the act of measurement doesn't make the system any more or less real. The reason why the moon orbiting earth seems a very classical system is only because we are accustomed to ask about it only those questions that we expect to have well determined, arbitrary function free answers. However, if we decided to ask the moon-orbiting-earth system a question similar to the question of where the electron is in an atom, we would probably want to ask something like "what is the street address of the moon?" That is a strange question to ask, given our classical expectations about the moon, but it is by all means an equivalent question to the question of where is the electron in an atom. It is a "gauge variant" question, because the only way we could answer it would be to define a projection operator that would project the moon to a specific street address. But so will your neighbour be able to project the same moon at his address, and hence all of a sudden, given this choice of observables, the moon will be highly delocalised at very many street addresses in a given geographical area. This would clearly show the "quantumness" of the moon-earth system. However, we know better than asking such questions about the moon, but it seems we don't know better about asking such questions of an electron in an atom. The result for the atom is a series of orbitals that we know how to determine and solve. The result for the moon are geographic areas that change daily, where the moon occupies a lot of different street addresses at the same time. In any case, the point of this argument is to make clear that the specific choice of an observable (even one on which several measurements won't agree due to various arbitrary functions that appear to be propagated along the solutions) defines the types of properties that we can reliably expect to determine. So, is the moon "quantum"? It seems that it totally depends on what questions we want to ask about it.
However, enough with the mind-experiments, and let me return to the formal proofs of my statements. 
If we think of gauge theory, we have to consider the propagation of gauge degrees of freedom together with the solutions of the differential equations. Because of that, it is expected that the equations of motion will not fully determine the dynamical variables for all times given only the initial conditions. This lack of determination is usually explained by means of a choice of a "frame" that could be different in the future and have a local nature. In any case, at least apparently, the form of the evolution in time may be different given different choices of such frames. In a gauge theory therefore, the general solution of the equations of motion contain arbitrary functions of time. 

A very illuminating discussion about the properties of gauge systems has been presented in ref. [23]. I will follow this common knowledge material from the mentioned textbook.
The fact that we have arbitrary functions of time in the solutions of the equations of motion makes it necessary to have dependent canonical variables. These dependencies are encoded via constraints and hence a gauge system will always appear as a constrained dynamical system. 
A gauge system is always a constraint system. Of course the reverse is not true, there exist other types of constraints on a Hamiltonian system that are not gauge in origin. However, it is important to note that the Hamilton Jacobi description of constraint systems is common to gauge systems and to quantum systems, and the particular way in which gauge systems emerge in the Hamilton Jacobi equation is, as will be shown further on, exactly the same as the way in which quantum mechanics is introduced. 
The classical dynamics of a system is that which makes the action 
\begin{equation}
S_{L}=\int_{t_{1}}^{t_{2}}L(q,\dot{q})dt
\end{equation}
stationary under variations of the variables $\delta q^{n}$ which vanish at the endpoints of the trajectory. 
The stationarity of the action is given by the Euler Lagrange equations which in their simplest form appear as
\begin{equation}
\frac{d}{dt}(\frac{\partial L}{\partial \dot{q}^{n}})-\frac{\partial L}{\partial q^{n}}=0
\end{equation}
which we can rewrite in order to make the transformation clearer
\begin{equation}
\ddot{q}^{n}\frac{\partial^{2}L}{\partial \dot{q}^{n'}\partial \dot{q}^{n}}=\frac{\partial L}{\partial q^{n}}-\dot{q}^{n'}\frac{\partial^{2}L}{\partial q^{n'}}\partial{\dot{q}^{n}}
\end{equation}
The accelerations can be determined by this equation in terms of velocities and position at a given time if the matrix 
\begin{equation}
\frac{\partial^{2}L}{\partial \dot{q}^{n'}\partial \dot{q}^{n}}
\end{equation}
can be inverted which leads us to the determinant condition 
\begin{equation}
det(\frac{\partial^{2}L}{\partial \dot{q}^{n'}\partial \dot{q}^{n}})\neq 0
\end{equation}
If this determinant is however zero, then we will not be able to determine the accelerations uniquely by positions and velocities and hence we will obtain solutions of the equations of motion that will depend on arbitrary functions of time. The non-invertibility of this function is therefore the starting point of gauge dynamics. The idea of gauge originates therefore from the idea that this function cannot be inverted. 
The absence of invertibility is associated to a one-to-many or multivariate function that will represent the dynamics. In this sense the equations of motion won't determine a unique solution and we won't have the possibility of finding one classical path. 

This is not surprising as in quantum mechanics we also do not have one given possible path, hence the Feynman path integral mechanism. However, there are other aspects we can show that are related to the fibre bundle quantisation prescription, in geometric quantisation, which I will here interpret as simply the result of the existence of a gauge bundle. In fact, there will be no direct distinction between a gauge fibre bundle and a quantum one given the symplectic structure. But first we have to make sure that the gauge bundle part of the formalism is clear. 
The Hamiltonian formalism is special by defining the canonical momenta 
\begin{equation}
p_{n}=\frac{\partial L}{\partial \dot{q}^{n}}
\end{equation}
and the non-invertibility of the previous function amounting to the vanishing of the determinant above is just the condition for the non-invertibility of the velocities as functions of the coordinates and momenta. We define the relations between the momenta and the positions as
\begin{equation}
\begin{array}{cc}
\phi_{m}(q,p)=0,& m=1,...,M
\end{array}
\end{equation}
These are primary constraints that are not derived from the equation of motion. 
The vanishing of the determinant above is the condition for non-invertibility of the velocities as functions of coordinates and momenta. This results in the fact that the momenta are not all independent, but are in fact related through $\phi_{m}$. When the momenta are expressed in the form of the equation 
\begin{equation}
p_{n}=\frac{\partial L}{\partial \dot{q}^{n}}
\end{equation}
the constraints $\phi_{m}$ are reduced to an identity. Therefore these constraints do not actually restrict the coordinates and velocities (or momenta), but instead they behave like generators of gauge transformations. 
We can imagine a $(q, \dot{q})$ space and consider the rank of the matrix 
\begin{equation}
\frac{\partial^{2}L}{\partial \dot{q}^{n}\partial \dot{q}^{n'}}
\end{equation}
to be constant throughout this space. Then our constraints $\phi_{m}$ define a submanifold smoothly embedded in the phase space. This submanifold is the primary constraint surface. 
If the rank of the above matrix is $N-M'$ then there are $M'$ independent equations among the constraints leading to a primary constraint surface which is a phase space submanifold of dimension $2N-M'$. The inverse transformation from $p$ to $\dot{q}$ is therefore multivalued resulting in a multivalued map 
$(q^{n},p_{n})\rightarrow (q^{n},\dot{q}^{n})$ that solves the equation of motion. We will have a map from a $2N$ manifold to a $2N-M'$ manifold. To make the transformation single valued we need to introduce extra parameters, at least $M'$ in number, that would discriminate the location of $\dot{q}$ on the inverse manifold. These parameters will appear as Lagrange multipliers in the Hamiltonian problem. 
If a smooth phase space function $G$ vanishes on the surface provided to us by the constraints then we can for given functions $g^{m}$ have $G=g^{m}\phi_{m}$. At the same time if $\lambda_{n}\delta q^{n}+\mu^{n}\delta p_{n}=0$ for arbitrary variations $\delta q^{n}$ and $\delta p_{n}$ tangent to the constraint surface we have
\begin{equation}
\begin{array}{c}
\lambda_{n}=u^{m}\frac{\partial \phi_{m}}{\partial q^{n}}\\
\\
\mu^{n}=u^{m}\frac{\partial \phi_{m}}{\partial p_{n}}\\
\end{array}
\end{equation}
for some $u^{m}$ with the equalities holding on the constraint surface. 
We can now introduce the Hamiltonian in a canonical way by 
\begin{equation}
H=\dot{q}^{n}p_{n}-L
\end{equation}
The hamiltonian is a function of position and velocities. The $\dot{q}$ enters H only through the combination $p(q,\dot{q})$ given by the hamiltonian equations of motion. Independent variations of positions and velocities lead to 
\begin{equation}
\begin{array}{c}
\delta H=\dot{q}^{n}\delta p_{n}+\delta \dot{q}^{n}p_{n}\frac{\partial L}{\partial \dot{q}^{n}}-\delta q^{n}\frac{\partial L}{\partial q^{n}}=\\
\\
=\dot{q}\delta p_{n}-\delta q^{n}\frac{\partial L}{\partial q^{n}}
\end{array}
\end{equation}
Here $\delta p_{n}$ is not an independent variation but instead is a linear combination of $\delta q$ and $\delta \dot{q}$. Therefore $\delta\dot{q}$ appears in the above equation only through that linear combination and not in another way. This makes $H$ a function of $p$ and $q$. However the $\delta p_{n}$ are not all independent but instead are restricted by the primary constraint $\phi_{m}=0$. 
The final result is that the canonical Hamiltonian is well defined only on the submanifold defined by the primary constraints but can be extended arbitrarily outside that manifold. We will use the same formalism if we wanted to make the replacement 
\begin{equation}
H\rightarrow H+c^{m}(q,p)\phi_{m}
\end{equation}
This is the construction of a bundle that lifts a base manifold via a fibre to an extended space in which we have arbitrary choices of frames. We consider a principal bundle $N\rightarrow Q$ with group $G$ and a connection $\alpha$ and we have a connection dependent projection $T^{*}N\rightarrow T^{*}Q$. The $G$-action on $N$ can be lifted to a Hamiltonian $G$-action on $T^{*}N$ with moment map $J:T^{*}N\rightarrow \mathbb{g}^{*}$. Let us consider a certain $\mu\in \mathbb{g}^{*}$ such that the associated co-adjoint orbit $\mathcal{O}_{\mu}$ is integral, then we have the isomorphism 
\begin{equation}
E=J^{-1}(\mathcal{O}_{\mu})/G\sim J^{-1}(\mu)/H
\end{equation}
leading to the Marsden-Weinstein reduced space, with $H$ the isotropy subgroup of $\mu$. With the Marsden-Weinstein reduction we construct a new symplectic manifold for a mechanical system with symmetry by a canonical action of a Lie group with or without the presence of a momentum map. With a basis given as $B=T^{*}Q$ then $E\rightarrow B$ becomes a fibre bundle with the fibre $F=\mathcal{O}_{\mu}$ and the transition function is induced by the $G$-action on $F$, preserving the canonical one-form. Adding a Kahler polarisation would produce the standard quantisation of the fibre $F$ and an irreducible representation space $\mathcal{Q}(F)$ of $G$ induced by a phase $U(1)$ representation of $H$. Not surprisingly these are also interpreted as the "internal" symmetries of the particle of charge $\mu$. We clearly cannot separate in a system with arbitrary functions or in a gauge system, the symmetry we obtain from the gauge group from the quantum "phase" symmetry. We may however decide not to introduce the symplectic structure and the Kahler polarisation, but that would yield a phase structure that would still contain what I mentioned above as non-separability or entanglement, however we would have to constrain the gauge degrees of freedom similarly later on to encode the correlation between non-realised degrees of freedom differently. I am not aware of other forms of encoding that, but I can imagine people have tried for example quaternion representations of quantum mechanics, etc. It is clear however that, no matter what the representation, non-determination of conjugate variables as well as of gauge degrees of freedom represented as arbitrary functions for incomplete integrals, lead to the same interpretation of a wavefunction (or quantum field). The Hamilton Jacobi equation with incomplete integrals and constraints is mapped directly into a quantum formulation with a complex wavefunction by expanding to a Kahler polarisation. While I agree it is more comfortable to work with a complex phase, its role to correlate in a non-trivial manner non-realised intermediate states is the same as what a gauge symmetry does.
The variation of the Hamiltonian can be re-written as 
\begin{equation}
(\frac{\partial H}{\partial q^{n}}+\frac{\partial L}{\partial q^{n}})\delta q^{n}+(\frac{\partial H}{\partial p_{n}}-\dot{q}^{n})\delta p_{n}=0
\end{equation}
and then 
\begin{equation}
\begin{array}{c}
\dot{q}^{n}=\frac{\partial H}{\partial p_{n}}+u^{m}\frac{\partial \phi_{m}}{\partial p_{n}}\\
-\frac{\partial L}{\partial q^{n}}|_{\dot{q}}=\frac{\partial H}{\partial q^{n}}|_{p}+u^{m}\frac{\partial \phi_{m}}{\partial q^{n}}\\
\end{array}
\end{equation}
From the first relation above we can extract the velocities $\dot{q}^{n}$ by using the momenta $p_{n}$ obeying $\phi_{m}=0$ and the extra parameters $u^{m}$. We can imagine those extra parameters as a choice of coordinates on the surface of the inverse images of $p_{n}$. 

Given the independent constraints, we also have the vectors $\frac{\partial \phi_{m}}{\partial p_{n}}$ independent on the surface $\phi_{m}=0$ (given some regularity conditions). 
Therefore no two different sets of $u$ can yield the same velocities.
We can therefore express $u$ as functions of the coordinates and velocities by solving 
\begin{equation}
\dot{q}^{n}=\frac{\partial H}{\partial p_{n}}(q,p(q,\dot{q}))+u^{m}(q,\dot{q})\frac{\partial \phi_{m}}{\partial p_{n}}(q,p(q,\dot{q}))
\end{equation}
If we define the Legendre transformation from $(q,\dot{q})$ space to the surface $\phi_{m}(q,p)=0$ in the $(q,p,u)$ space by using 
\begin{equation}
\begin{array}{c}
q^{n}=q^{n}\\
\\
p_{n}=\frac{\partial L}{\partial \dot{q}^{n}}(q,\dot{q})\\
\\
u^{m}=u^{m}(q,\dot{q})
\\
\end{array}
\end{equation}
we see that this transformation between spaces of the same dimensionality is invertible because
\begin{equation}
\begin{array}{c}
q^{n}=q^{n}\\
\\
\dot{q}^{n}=\frac{\partial H}{\partial p_{n}}+u^{m}\frac{\partial \phi_{m}}{\partial p_{n}}\\
\\
\phi_{m}(q,p)=0\\
\end{array}
\end{equation}
and therefore invertibility of the Legendre transformation when $det(\frac{\partial^{2}L}{\partial \dot{q}^{n}\partial \dot{q}^{n'}})=0$ can be regained if we add extra variables. 
We can understand the $u^{m}$ functions as Lagrange multipliers for our gauge constraints.
We can therefore re-write the original Lagrangian equations in the Hamiltonian form as
\begin{equation}
\begin{array}{c}
\dot{q}^{n}=\frac{\partial H}{\partial p_{n}}+u^{m}\frac{\partial \phi_{m}}{\partial p_{n}}\\
\\
\dot{p}_{n}=-\frac{\partial H}{\partial q^{n}}-u^{m}\frac{\partial \phi_{m}}{\partial q^{n}}\\
\\
\phi_{m}(q,p)=0\\
\\
\end{array}
\end{equation}
The Hamilton equations can then be recovered via the variational principle from 
\begin{equation}
\delta \int_{t_{1}}^{t_{2}}(\dot{q}^{n}p_{n}-H-u^{m}\phi_{m})=0
\end{equation}
for arbitrary variations $\delta q^{n}$, $\delta p_{n}$, $\delta u_{m}$ subject to $\delta q^{n}(t_{1})=\delta q^{n}(t_{2})=0$. The new variables making the Legendre transform invertible are now Lagrange multipliers for the primary constraints. 

The theory is clearly invariant to $H\rightarrow H+c^{m}\phi_{m}$ as this can be obtained as a shift in the Lagrange multipliers $u^{m}\rightarrow u^{m}+c^{m}$. We can therefore re-write the variational principle in an equivalent form with fewer variables but where our constraints are solved
\begin{equation}
\delta \int_{t_{1}}^{t_{2}}(\dot{q}^{n}p_{n}-H)dt=0
\end{equation}
with the constraints 
\begin{equation}
\begin{array}{c}
\phi_{m}=0\\
\delta \phi_{m}=0\\
\end{array}
\end{equation}

This leads to an equation of motion given the first variational principle (see above) of the form 
\begin{equation}
\dot{F}=[F,H]+u^{m}[F,\phi_{m}]
\end{equation}
where $F$ is an arbitrary function of the canonical variables and the Poisson bracket is 
\begin{equation}
[F,G]=\frac{\partial F}{\partial q^{i}}\frac{\partial{G}}{\partial p_{i}}-\frac{\partial{F}}{\partial p_{i}}\frac{\partial G}{\partial q^{i}}
\end{equation}
If we require that the primary constraints are preserved in time, we can take $F=\phi_{m}$ and demand $\dot{\phi}_{m}=0$. This gives us the condition
\begin{equation}
[\phi_{m},H]+u^{m'}[\phi_{m},\phi_{m'}]=0
\end{equation}
This can become completely independent of $u$ or it may impose an additional restriction on $u$. In the first case, if the remaining relation between coordinates and momenta is independent of the primary constraints, we call it a secondary constraint. Secondary constraints differ from primary ones because primary constraints are solely the results of $p_{n}=\frac{\partial L}{\partial \dot{q}^{n}}$ that define the momentum variables, while for secondary constraints one must also use the equations of motion. 
If a secondary constraint appears, for example $X(q,p)=0$ we must impose another condition 
\begin{equation}
[X,H]+u^{m}[X,\phi_{m}]=0
\end{equation}
With this we have to check again if it implies new secondary constraints or it only restricts the $u$, etc. After this tower of checks is finished we remain with a set of secondary constraints 
\begin{equation}
\begin{array}{cc}
\phi_{k}=0, & k=M+1, ..., M+K\\
\end{array}
\end{equation}
with $K$ the total number of secondary constraints. A better notation would be that of weak equality, and therefore we may write that the constraint, for a general $j=1, ..., M+K$ is $\phi_{j}\sim 0$. This means that $\phi_{j}$ is numerically restricted to zero but does not identically vanish throughout the whole phase space. It may therefore have non-zero Poisson brackets with the canonical variables. In general, two functions $F$ and $G$ that are identical on the submanifold defined by the constraints $\phi_{j}\sim 0$, are called weakly equal $(F\sim G)$. An equation that holds throughout the phase space and not just on the constraint manifold is called strong. We have therefore 
\begin{equation}
F\sim G \leftrightarrow F-G=c^{j}(q,p)\phi_{j}
\end{equation}
Given a complete set of constraints there will be restrictions on the Lagrange multipliers $u^{m}$ of the form 
\begin{equation}
[\phi_{j}, H]+u^{m}[\phi_{j},\phi_{m}]\sim 0
\end{equation}
with $m=1,...,M$ and $j=1,...,J$. This appears as a set of $J$ non-homogeneous linear equations with $M\leq J$ unknowns $u^{m}$. The coefficients are functions of $p$ and $q$. The general solution of such a system is of the form 
\begin{equation}
u^{m}=U^{m}+V^{m}
\end{equation}
where $U^{m}$ is the particular solution of the inhomogeneous equation and $V^{m}$ is the general solution of the associated homogeneous system
\begin{equation}
V^{m}[\phi_{j}, \phi_{m}]\sim 0
\end{equation}
$V^{m}$ takes most generally a linear combination of independent solutions $V_{a}^{m}$, $a=1,...,A$. The number $A$ of independent solutions is the same for all $q$ and $p$ satisfying the constraints because we assumed the matrix $[\phi_{j},\phi_{m}]$ to be of constant rank where the constraints are satisfied. The general solution of the original equation will therefore be 
\begin{equation}
u^{m}\sim U^{m}+v^{a}V_{a}^{m}
\end{equation}
in which the coefficients $v^{a}$ are arbitrary. We have therefore written the general solution in a form in which we separated the completely arbitrary functions from those that are fixed by the consistency conditions resulting from the demand that the constraints are preserved in time. The equations $\phi_{j}=0$ are said to be reducible if the constraints are not all independent. When all the constraints are independent we say the equation is irreducible. We may decide to eliminate the dependent constraints, but when doing this one may abandon manifest invariance to some symmetries, or the process may even be impossible to perform globally, due to topological obstructions. We re-write now $\dot{F}=[F,H]+u^{m}[F, \phi_{m}]$ in the form 
\begin{equation}
\dot{F}\sim [F, H'+v^{a}\phi_{a}]
\end{equation}
with $H'=H+U^{m}\phi_{m}$ and $\phi_{a}=V_{a}^{m}\phi_{m}$ where $H_{T}=H'+v^{a}\phi_{a}$ is the total Hamiltonian. Therefore 
\begin{equation}
\dot{F}\sim [F, H_{T}]
\end{equation}
will contain $A$ arbitrary functions $v^{a}$ and represent the original Lagrangian equations of motion. 
We call a function $F(q,p)$ of first class, if its Poisson bracket with every constraint vanishes weakly
\begin{equation}
[F,\phi_{j}]\sim 0, ;\ j=1,...,J
\end{equation}
The Poisson bracket of two first class functions will remain a first class function. Both $H'$ and $\phi_{a}$ are first class. $\phi_{a}$ is a complete set of first class primary constraints and hence any first class primary constraint is a linear combination of $\phi_{a}$ with coefficients that are functions of $q$ and $p$ and module squares of second class constraints. The total Hamiltonian is the sum of the first class Hamiltonian $H'$ and the first class primary constraints multiplied by arbitrary coefficients. The splitting of the total Hamiltonian into $H'$ and $v^{a}\phi_{a}$ is not unique because $U^{m}$ can be any solution of the inhomogeneous equation. Therefore shifting $v^{a}$ we can introduce into $H'$ any linear combination of $\phi_{a}$ without altering the total Hamiltonian. The fact that we have arbitrary functions $v^{a}$ in the total Hamiltonian means that not all the $q$'s and $p$'s are observable. The physical state may be uniquely described once a set of $q$'s and $p$'s is given but there is more than one set of values for the canonical variables that represent a given physical state. 
Let us start with a physical system determined at time $t_{1}$ and propagate it in time. If we provided a canonical set of variables at that initial time, we would expect that propagating the system in time will keep it fully determined at any further time. However, the existence of arbitrary functions $v^{a}$ that can indeed be chosen independently at any subsequent time alters this determined evolution. In fact any further evolution of the system will depend on the choice of the values for $v_{a}$ in any time point between $t_{1}\leq t\leq t_{2}$. If we write $t_{2}=t_{1}+\delta t$ then the difference of the dynamical variable $F$ at time $t_{2}$ corresponding to two different choices $v^{a}$ and $\tilde{v}^{a}$ of the arbitrary functions at time $t_{1}$ is
\begin{equation}
\delta F=\delta v^{a}[F,\phi_{a}]
\end{equation}
given $\delta v^{a}=(v^{a}-\tilde{v}^{a})\delta t$. This shows that indeed first class primary constraints do generate gauge transformations. 
With this we can in principle repeat the path integral quantisation prescription for gauge degrees of freedom. Most of the time we want the exact opposite. Because we first attempt to quantise a theory and hence make it dependent on unrealised, equivalent, (given the experimental context) intermediary states with usually several possible outcomes (which we can distinguish only after the realisation of the experiment, but not within the experiment), and by doing this we need to sum/integrate over those unrealised intermediary states. It also happens that when we do this in a purely formal way, we also integrate over the gauge degrees of freedom if our theory is a gauge theory. This integration is to some extent spurious and hence we needed to restrict it so that each gauge equivalence class is counted only once, by means of its chosen representative. 
Therefore creating a Feynman path integral formulation for gauge degrees of freedom would look at a first glance absurd. But the absurdity only comes from our assumption that the gauge degrees of freedom are identical and unphysical. Non-linearity in non-abelian gauge theories however leads to persistent Gribov ambiguities, and to situations in which the Gribov copies (clearly "gauge artefacts" if looked upon from the traditional viewpoint) play an important role and describe actual features of the theory. Of course, those features result in a multiple counting, and we could look at them as a spurious repetition, if it wasn't for the fact that it makes us realise there are other features for which this doubling is an indicator, that are totally not spurious. 

From this point of view, the dynamics of the system is not altered. This raises a question: if we are to perform a path integral quantisation prescription a la Feynman, we say that we integrate over indistinguishable paths, from the perspective of the given experiment, that do however have clearly distinct canonical variables and appear to us as distinguishable. After all, in the double slit experiment we say we can at least in principle attach a detector to one of the slits, making the "which path" question clearly answerable. Nevertheless, being able to change an experimental setup doesn't change the inability to distinguish the paths in the given setup. The paths are, as far as the experiment is concerned, indistinguishable. The question arises, is there any difference between that situation and the situation in which we deal with gauge modifications that are claimed to be "unphysical"?
The point I am trying to make in this article is that the answer to this question is "essentially not". 
In a gauge context, the different forms that a gauge variant observable may take due to different choices of gauge, make it look like different observables, and in fact it really is so, to observers picking different frame (or gauge) choices. We can however identify those observables as "indistinguishable" from the perspective of some "absolute" experiment designed to be invariant to choices of gauge, but that again changes the experiment, eliminating the gauge dependent observables or fixing the gauge to one and only one choice. We basically define an equivalence (or cohomology) class and pick representatives of that class. The basis of BRST quantisation is exactly that and is assumed to be well known. If we think of observables as "questions to ask a physical system", then we basically change the questions and hence the design of the experiment so that we can get unambiguous answers. But it seems like the desire for unambiguous answers is a "human, all too human" desire...
We are getting again confused by what we could do or ought to do as opposed to what we do. We can indeed mathematically distinguish gauge equivalent trajectories, although the physics of those distinct trajectories remains the same in terms of the directly measurable effects. As we showed, there are a series of parameters/variables we can use to distinguish them and there is even a manifold over which we can integrate them. In various discussions of the Gribov ambiguity it was also wrongly assumed that we have to avoid integrating over Gribov copies in order to obtain confinement. In fact it turned out that taking into account those Gribov copies as actual topological features of the theory, we are better off at describing confinement. In this situation our view about what we declared "physical" and what not also changed a bit, receiving a historical side-note. Small gauge transformations were seen as "un-physical" while large gauge transformations were seen as having some physical effects given some topological (homotopy) structure of the gauge group. In any case, one historical side-note after another, we had to slowly update what we mean by "physical" and "un-physical" in this context. Aside of that problem that will be discussed in a different article, the distinction between "physical" and "unphysical" should be more blurred than we may want to admit in our current theories. It has been suggested that the distinction between a quantum path integral approach and a gauge theory approach is that in a quantum approach, for example, we are actually able to distinguish between alternative paths "physically", while in a gauge theory we are not, as the different "trajectories" in a gauge equivalence class are "truly" indistinguishable. My question is: are we really? If the double slit experiment is maintained in the way it has been designed, therefore, given the pre-set context, in fact, we are not able to practically distinguish the paths the electron (or photon, etc.) could take. That is why we obtain the interference pattern in quantum mechanics. Those paths are as "unphysical" (or as "physical") as the ones we see in the gauge choice of arbitrary functions. In both cases we make a choice, in one case the choice is the experimental setup, by which we evolve the experiment, in the second is the choice of a gauge, which we can change accordingly. Again, it is more of a physics folklore to call gauge related configurations as "unphysical", because we never see them, but we also never see the "distinguished" paths in a double slit experiment, if the experiment doesn't involve detectors on at least one slit. We can literally, and mathematically, repeat the gauge equivalence calculation for calculations of the path integral in the double slit experiment, with similar results, the only modification being that once we call the different trajectories as belonging to a "gauge class" and another time we call them "indistinguishable" in terms of the given experimental setup. 
To be more explicit, there is a distinction in the way in which we quantise gauge theories and non-gauge theories. In non-gauge theories, we do indeed promote observables to operators and take into account unrealised intermediate states in calculating probability amplitudes. If we do the same thing in gauge theories we end up with well known divergences due to over-integration over "spurious gauge groups". This happens precisely because the two methods, of describing gauge theories and quantisation are technically the same, the differences appearing only in our interpretation of the two. Once we consistently fix the gauge or implement some BRST or BV quantisation prescriptions, we reduce gauge ambiguities, but we do keep the original "physical" ambiguities related to, for example, "which path" questions. This is done precisely because the same mathematical tools that allow us to add gauge structure are also those that allow us to add "which path" ambiguities which we consider necessary to calculate amplitudes properly.
\par This approach to gauge symmetric theories is somehow different from the fibre bundle approach that was used previously but it has the advantage that it can be linked easier with quantum mechanics, by means of the Hamilton Jacobi theory. In a gauge theory case, the Hamilton Jacobi theory can be brought to a form equivalent to the one we use in the fibre bundle gauge approach showing that the quantum case basically just expands on non-determinability of constants of motion and their undetermined conjugates. 
It is probably important to mention that this article has nothing to do with the emergence of quantum mechanics from any classical theory, in fact I do not believe such an emergence exists. The world seems quantum, the only difference is that at some levels (for example higher length scales) the quantum effects take some different forms that we still need to learn how to recognise. However, the world is certainly not classical. One way of connecting gauge to quantum is by means of the Hamilton Jacobi equation. It is worth mentioning what the idea behind this association is: first, I do not assume there exists a duality between classical and quantum mechanics, and therefore I do not wish to "derive" quantum mechanics from a classical world. This must be stated clearly because I am actually starting with a theory that is itself quantum, and I want to show that the way we describe it in terms of a dynamical system is equivalent to the way in which gauge is being introduced in a Hamilton Jacobi theory. That does not mean that I obtain quantum mechanics from a classical Hamilton Jacobi theory. It means that the gauge symmetry is also a quantum phenomenon. Because of this, many theories that are considered classical nowadays, like classical Maxwell equations, or even classical general relativity do have hidden quantum components. 
It also is important to say that the introduction of a gauge in the Hamilton Jacobi equation already makes the theory at least in part quantum, in the sense of allowing certain constraints that would make it afterwards quantum. Those constraints are not essential now. What is essential is to note that a gauge theory, even one that is considered classical, is in fact more quantum than a non-gauge classical theory. This is why a so called classical electromagnetic wave allows for entanglement and global non-separable structures, as has been shown in reference [13]. 
I am going to review for a while the basics of the unconstrained and constrained Hamilton Jacobi theory as well as the formulation in terms of complete and incomplete integrals. I will also show that the introduction of gauge arbitrary functions allows for various constraint problems that become equivalent to undetermined incomplete integral descriptions of Hamilton Jacobi. These lead in the case in which all variables obtained are non-determined to standard quantum mechanics. Again, the transition from classical to quantum is the same as done in the usual prescription of quantisation, the appearance of quantum mechanics in itself is not the issue here, instead, the issue is the role of gauge as a form of quantum representation. 
So, the equations of motion define canonical transformations between the coordinates $q^{i}$ and the momenta $p_{i}$ at time $t$ and the coordinates $q_{0}^{i}$ and momenta $p_{i}^{0}$ at time $t_{0}$. Let there be the canonical coordinates $(\alpha^{i},\beta_{i})$ obtained from $(q^{i}_{0}, p_{i}^{0})$ by a time independent canonical transformation. The transformation is also canonical and is obtained by means of a generating function $S(q^{i},\alpha^{i},t)$. We obtain 
\begin{equation}
\begin{array}{c}
p_{i}=\frac{\partial S}{\partial q^{i}}\\
\\
\beta_{j}=-\frac{\partial S}{\partial \alpha^{j}}\\
\\
\end{array}
\end{equation}
and 
\begin{equation}
det(\frac{\partial^{2}S}{\partial \alpha^{j}\partial q^{i}})\neq 0
\end{equation}
The variables $\alpha^{i}$ and $\beta_{i}$ are constants of motion. Hence the Hamiltonian $\bar{H}(\alpha^{i},\beta_{i})=H+\frac{\partial S}{\partial t}$ on which the evolution of the $\alpha^{i}$ and $\beta_{i}$ relay upon can be taken to zero. This will lead to the usual Hamilton Jacobi equations
\begin{equation}
\frac{\partial S}{\partial t}+H(q^{i},\frac{\partial S}{\partial q^{i}})=0
\end{equation}
The solution of this equation is the generating function for our canonical transformation $S(q^{i},\alpha^{i})$ and it depends on the $n$ variables $\alpha^{i}$ such that the determinant condition holds. This generating function is called a complete integral and its knowledge leads to the construction of the general solution of the equations of motion by simple substitution
\begin{equation}
\beta_{j}=-\frac{\partial S}{\partial \alpha^{j}} \leftrightarrow q^{i}=q^{i}(\alpha^{j},\beta_{k},t)
\end{equation}
The Hamilton principal function $W$ is that for which the time independent canonical variables are the initial data. In the case of unconstrained systems we discuss the situation in which the solution of the Hamilton Jacobi equation depends on fewer integration constants. In that case we call the solution incomplete integral and we obtain it by fixing $m$ of the $\alpha$ to be equal to definite values in the complete integral. If this happens then we denote the rest $\alpha^{A}$ as being the unspecified constants of motion, while the fixed ones we call $\alpha_{a}$. We may set those values to zero and then the incomplete integral will stop depending on a series of such constants $\alpha_{a}$. Because of this lack of dependence of $S$ on $\alpha_{a}$ the conjugate variables $\beta^{a}=-\frac{\partial S}{\partial \alpha_{a}}$ becomes unknown. The result of this is that by using the set of equations
\begin{equation}
\begin{array}{c}
p_{i}=\frac{\partial S}{\partial q^{i}}\\
\\
\beta_{A}=-\frac{\partial S}{\partial \alpha^{A}}\\
\\
rank(\frac{\partial ^{2}S}{\partial \alpha^{A}\partial q^{i}})=n-m\\
\\
\end{array}
\end{equation}
with both $(\alpha^{A}, \beta_{A})$ given, a complete integral $S(q^{i},\alpha^{A},t)$ cannot determine a unique solution of the equations of motion. This means that the constants of motion $\alpha^{A}$ and $\beta_{A}$ with $\alpha_{a}=0$ do not characterise a single classical trajectory. They do characterise all trajectories that differ in value by the unknown conjugate $\beta^{a}$ of $\alpha_{a}$. This means that if we know $(q^{i},p_{i})$ to be a solution of the above equations based on an incomplete integral at time $t$ then with a solution $(q^{i}+\delta q^{i}, p_{i}+\delta p_{i})$ at time $t+\delta t$ we would have 
\begin{equation}
\begin{array}{c}
\delta q^{i}=\frac{\partial H}{\partial p_{i}}(q,p)\delta t\\
\\
\delta p_{i}=-\frac{\partial H}{\partial q^{i}}(q,p)\delta t\\
\\
\end{array}
\end{equation}
if the two solutions happen to give the same values of the conjugate momenta $\beta^{a}$ and therefore lie on the same classical trajectory. This doesn't have to happen. In fact this may happen in a probabilistic sense which leads to the construction of intermediate states and the Feynman path integral approach and even simple quantum mechanics in which the probability of non-realised states intermediate states must also be included as probability amplitudes, and for which we define then the Born rule. In any case, if there are different values for $\beta^{a}$ for the solutions at the two different time-steps we obtain 
\begin{equation}
\begin{array}{c}
\delta q^{i}=[\frac{\partial H}{\partial p_{i}}(q,p)+\lambda^{a}\frac{\partial \alpha_{a}}{\partial p_{i}}(q,p)]\delta t\\
\\
\delta p_{i}=[-\frac{\partial H}{\partial q^{i}}(q,p)-\lambda^{a}\frac{\partial \alpha_{a}}{\partial q^{i}}(q,p)]\delta t\\
\\
\end{array}
\end{equation}
with some values for $\lambda^{a}$ and the condition that $\alpha_{a}(q,p)=0$. If $\lambda^{a}$ vanishes then we obtain a classical trajectory. The limit case in which the solution $S(q^{i},t)$ involves no integration constant at all is that in which any two solutions have the same values of a complete set of commuting conserved observables but have different conjugates. This is the case relevant for quantum mechanics. 
In the case of constrained systems we have exactly the same situation as with incomplete solutions. We just distinguish them in an arbitrary fashion. 
We may simplify things if we identify $\alpha_{a}$ with an abelian representation of the constraint surface $G_{a}=0$. The conjugate variables $\beta^{a}$ are then pure gauge while the other variables $\alpha^{A}$ and their conjugates which commute with $\alpha_{a}$ form a complete set of gauge invariant functions. They can be considered canonical coordinates on the reduced phase space or be associated with a complete set of observables. The generating function $S(q^{i},\alpha^{A},\alpha_{a},t)$ defines a canonical transformation 
\begin{equation}
(q^{i},p_{i})\rightarrow \alpha^{A},\beta_{A},\alpha_{a},\beta^{a}
\end{equation}
such that the constraints become $\alpha_{a}=0$. Then the generating function will obey
\begin{equation}
\begin{array}{c}
G_{a}(q^{i},\frac{\partial S}{\partial q^{i}})=0\\
\\
\frac{\partial S}{\partial t}+H_{0}(q^{i},\frac{\partial S}{\partial q^{i}})=0\\
\\
rank(\frac{\partial^{2}S}{\partial \alpha^{A}\partial q^{i}})=n-m\\
\end{array}
\end{equation}
The information on $\frac{\partial S}{\partial \alpha_{a}}=-\beta^{a}$ is lost and hence the momentum type variables become arbitrary. These are now the Hamilton Jacobi equations for a constraint system. They already are gauge, and in the interpretation of this article they are also quantum. Which makes us think whether quantum is basically gauge. In any case, if $(q^{i}(t),p_{i}(t))$ is a solution for each $t$ for
\begin{equation}
\begin{array}{c}
p_{i}=\frac{\partial S}{\partial q^{i}}\\
\\
\beta_{A}=-\frac{\partial S}{\partial \alpha^{A}}\\
\\
\end{array}
\end{equation}
we obtain 
\begin{equation}
\begin{array}{c}
\dot{q}^{i}=[q^{i},H_{0}]+\lambda^{a}[q^{i},G_{a}]\\
\\
\dot{p}_{i}=[p_{i},H_{0}]+\lambda^{a}[p_{i},G_{a}]\\
\\
G_{a}(q,p)=0\\
\end{array}
\end{equation}

this means that $(q^{i}(t),p_{i}(t))$ is a solution of the equations of motion for some choice of the multipliers $\lambda^{a}$. It is interesting to note that the Hamilton Jacobi function $S$ contains already all solutions of the equations, even those related by a gauge transformation. Those all appear as a phase in the path integral formulation. 
The distinction between those solutions that are "physically relevant" but not realised and those that are "not physically relevant" and still not realised is here being made, and when we start the quantisation of gauge theories, many of them were already counted for in the process of constructing the gauge theory and this is why we require some gauge fixing. We just thought of them as redundant, but this redundancy comes from the fact that gauge theories and quantum theories have a common origin and are to the extent of their working principles the same thing. 
What gauge we fix and in what way is equivalent to what observables we choose to be in the complete set of commuting observables, with the observation that no absolute choice for those exists, as there is no absolute choice of a gauge. Can we define a probabilistic description of gauge choices? Yes, in fact it is possible, if we assign probability or probability distributions over the gauge space itself. It would require some careful consideration of the normalisation conditions, but there is no reason not to be able to do it, having ultimately an arbitrary function that is involved. As before, solutions of $S(q^{i},t)$ that depend on no integration constants whatsoever are in perfect analogy to quantum mechanics. The question would be why some of the fixed integration constants vanish? Is it because of a choice or because of the fact that gauge invariance sets them to zero? From the perspective of the theory there is no reason one should distinguish the situation in which the conjugate momenta $\beta^{a}$ are unknown (as is the case in quantum mechanics) or are pure gauge variables that are assumed to be "undetermined in principle". This is the same distinction between quantum mechanics and gauge theories. The non-determination in quantum mechanics is one that appears by choice or by nature? It could very well be that the reason doesn't truly matter, as long as they are fundamentally undetermined, but then there is no clear distinction between quantum theories and gauge theories. Which is also the claim of this article. 
If the constraints $G_{a}$ are linear and homogeneous in the momenta, 
\begin{equation}
G_{a}=a_{a}^{i}(q)p_{i}
\end{equation}
we obtain 
\begin{equation}
a_{a}^{i}(q)\frac{\partial S}{\partial q^{i}}=0
\end{equation}
and that makes the evolution equations for $q$ closed (i.e. involving just other $q$'s)
\begin{equation}
q^{i}\rightarrow q^{i}+\epsilon^{a}a_{a}^{i}(q)
\end{equation}
meaning we defined internal gauge symmetries. 
This means
\begin{equation}
S(q+\delta_{\epsilon}q)-S(q)=0
\end{equation}
which means that the gauge invariance of $S$ is $\delta_{\epsilon}S=0$ if $S$ transforms like a scalar
\begin{equation}
\delta_{\epsilon}S=S(q+\delta_{\epsilon}q)-S(q)
\end{equation}
If $S$ transforms inhomogeneously for example because 
\begin{equation}
G_{a}=p_{a}-\frac{\partial V}{\partial q^{a}}(q)
\end{equation}
we get
\begin{equation}
\delta_{\epsilon}S=S(q+\delta_{\epsilon}q)-S(q)-[V(q+\delta_{\epsilon}q)-V(q)]
\end{equation}
When the constraints are non-linear in the momenta, the notion of gauge invariance cannot directly be inferred from the constraint and Hamilton-Jaconi equations above. However, the concept naturally exists in quantum mechanics where we get the linear equations $G_{a}\psi=0$ for a wavefunction $\psi$. If $\psi$ does not transform like a scalar this is of course not a problem anymore. In any case, this shows the strong conceptual connection between a quantum theory and a gauge theory. The generalisation that we do in quantum mechanics is really of the same type, with the distinction that the Poisson bracket would be meaningless or lead to contradictions and hence we have to replace it with operators and commutation relations. But the quantum nature is constructed in the same way. 
On the other side, in quantum mechanics a similar problem appears: if we work in the reduced phase space that only allows for gauge independent observables to be those realised as quantum mechanical operators and every state in the Hilbert space has to be physical then by eliminating the gauge degrees of freedom and finding only a complete set of gauge invariant observables, we loose manifest invariance under symmetries like Lorentz group. If we have a theory that only is represented in terms of a complete set of independent observables like, in the case of a free relativistic particle, described by the constraint $p^{2}+m^{2}=0$, we would have a complete set of independent observables given by 
\begin{equation}
\begin{array}{cc}
p^{i},\; & x^{i}-(p^{i}x^{0}/\sqrt{m^{2}+p^{j}p_{j}})
\end{array}
\end{equation}
which would not be in any linear representation of the Lorentz group which would make any unconstrained theory expressed in this way not (linearly) Lorentz invariant. At the same time, if we decided to eliminate gauge degrees of freedom in a field theoretic context we would not have locality in spacetime. Moreover, if we used the Hamiltonian in terms of the independent degrees of freedom we would obtain a very complicated expression which would be impossible to quantise to begin with. While there exist quantisation descriptions for all types of gauge theories, from BRST to BV or further to various types of closures of the gauge algebra, this is not the point here. The point is, the gauge degrees of freedom emerge in the same way as quantum degrees of freedom and in this sense the two concepts are dual. The linearity condition in the representation of the Lorentz invariance would not be possible with only "physical" degrees of freedom, which would also conflict with a linear realisation of relativistic quantum mechanics. In any case, there seems to be a strong connection between gauge and quantum. 
Recent research has given the Hamilton Jacobi theory a geometric interpretation using fibre bundles. Using that interpretation for the Hamilton Jacobi equation we arrive at the same conclusion as above, that if gauge invariance is included, the separability of trajectories cannot be maintained or even unambiguously defined, leaving basically gauge connections to be fundamentally entangled structures, as we move from one fibre to the next. 
\section{The geometric representation of Hamilton Jacobi theory and its quantum / gauge analogue}
We have seen in the previous chapter how quantum mechanics appears as the geometric construction of Hamilton Jacobi in the case in which we have incomplete integrals and we make a series of constants of motion be eliminated from $S$ leading to undetermined conjugate momenta. To make clear why the fibre bundle construction is equivalent to the entanglement feature of the fibre bundle it is desirable to form a geometric representation of the Hamilton Jacobi theory in terms of fibre bundles. Such a construction actually exists [14, 15, 16]. The reformulation of that construction in terms of incomplete integrals will result in a fibre bundle approach in which the non-separability will originate from the gauge arbitrary functions included, and that will basically be quantum (entanglement). Therefore the way of thinking is: we start with a Hamilton Jacobi theory, we re-define it in terms of a geometric fibre bundle approach, we then implement the gauge invariance strategy into it by eliminating constants of motions and making the conjugate variables undetermined. The resulting geometry is one in which the non-separability due to entanglement emerges. This means that gauge theories must be fundamentally quantum even when they are being considered classical. The next step would be to promote the observables to operators in order to restore the quantum commutator brackets. That part is essential for quantum mechanics, of course, and a classical theory with gauge written in the previous way would be inconsistent, but then a classical theory is fundamentally inconsistent. What I am trying to show is a link between a gauge theory and a quantum theory, and not the emergence of quantum from classical. We construct quantum mechanics by avoiding a series of inconsistencies in classical mechanics, but an important aspect of quantum mechanics is also the existence of gauge. 
Looking at the Hamilton-Jacobi theory, the canonical transformations are associated with a foliation in the phase space of the system, which we can then show to be a cotangent bundle $T^{*}Q$ of the configuration manifold $Q$. This foliation is invariant to the dynamics, transversal to the fibres of the cotangent bundle, and the restriction of the dynamical vector field in $T^{*}Q$ to each leaf $S_{\lambda}$ of this foliation projects onto a vector field $X_{\lambda}$ on $Q$. If the integral curves of these vector fields are one-to-one related, the complete set of dynamical trajectories are the integral curves of the complete family $\{X_{\lambda}\}$ of all the vector fields in the base. The geometric Hamilton Jacobi problem then means just finding this foliation and the vector fields $\{X_{\lambda}\}$. 
Let us consider the Hamiltonian system defined by $(T^{*}Q,\omega, H)$ and the bundle $\pi_{Q}:T^{*}Q\rightarrow Q$ represents the phase space of the dynamical system and $Q$ is the configuration space. $\omega=-d\theta\in\Omega^{2}(T^{*}Q)$ is the symplectic form in $T^{*}Q$ and $H\in C^{\infty}(T^{*}Q)$ is the Hamiltonian function. The dynamical trajectories would be integral curves $\sigma:I\subset \mathbb{R}\rightarrow T^{*}Q$ of the Hamiltonian vector field $Z_{H}\in X(T^{*}Q)$ associated with H, which is the solution of the Hamiltonian equation 
\begin{equation}
i(Z_{H})\omega = dH
\end{equation}
where $\Omega^{k}(T^{*}Q)$ and $X(T^{*}Q)$ are the sets of differentiable $k$ forms and vector fields in $T^{*}Q$ and $i(Z_{H})\omega$ denotes the inner contraction of $Z_{H}$ and $\omega$. If we choose natural coordinates $(q^{i},p_{i})$ of $T^{*}Q$ we obtain $\omega=dq^{i}\wedge dp_{i}$ and the curves $(q^{i}(t),p_{i}(t))$ are solutions of the Hamilton equations
\begin{equation}
\begin{array}{cc}
\frac{d q^{i}}{dt}=\frac{\partial H}{\partial p_{i}}(q(t),p(t)), \;& \frac{dp_{i}}{dt}=-\frac{\partial H}{\partial q^{i}}(q(t),p(t))
\end{array}
\end{equation}
Therefore if we define a Hamiltonian Hamiltin-Jacobi problem for a Hamiltonian system characterised by $(T^{*}Q, \omega, H)$, we are interested in finding a vector field $X\in \mathcal{X}(Q)$ and an associated 1-form $\alpha \in \Omega^{1}(Q)$ such that if we have an integral curve of $X$, $\gamma:\mathbb{R}\rightarrow Q$ then $\alpha\circ \gamma :\mathbb{R}\rightarrow T^{*}Q$ is an integral curve of $Z_{H}$ i.e. if $X\circ \gamma=\dot{\gamma}$ then $\dot{\overline{\alpha\circ \gamma}}=Z_{H}\circ(\alpha\circ \gamma)$. The couple $(X,\alpha)$ is then a solution of the generalised Hamiltonian Hamilton Jacobi problem. 

We consider $\Lambda \subset \mathbb{R}^{n}$ and a family of solutions $\{\alpha_{\lambda}; \lambda\in \Lambda\}$, depending on $n$ parameters $\lambda=(\lambda_{1},\lambda_{2},...,\lambda_{n})$. This is a complete solution of the Hamilton Jacobi problem if the map 
\begin{equation}
\Phi: Q\times \Lambda\rightarrow T^{*}Q
\end{equation}
is a local diffeomorphism, 
leading to 
\begin{equation}
(q,\lambda)\rightarrow \alpha_{\lambda}(q)
\end{equation}
Given a complete solution $\{\alpha_{\lambda};\lambda\in \Lambda\}$ as $d\alpha_{\lambda}=0$, $\forall \lambda\in \Lambda$ we have a family of functions $\{S_{\lambda}\}$ defined on the neighbourhoods $U_{\lambda}\subset Q$ of every point such that $\alpha=dS_{\lambda}$ and we have a local generating function of the complete solution $\{\alpha_{\lambda};\lambda\in \Lambda\}$
\begin{equation}
S:\cap U_{\lambda}\times \Lambda\subset Q\times \Lambda \rightarrow \mathbb{R}
\end{equation}
where we have
\begin{equation}
(q,\lambda)\rightarrow S_{\lambda}(q)
\end{equation}
A complete solution defines a Lagrangian foliation in $T^{*}Q$ which is transverse to the fibres and such that $Z_{H}$ is tangent to the leaves. The functions that locally define this foliation are the components of a map 
\begin{equation}
F: T^{*}Q\xrightarrow{\Phi^{-1}}Q\times \Lambda\xrightarrow{pr_{2}}\Lambda\subset \mathbb{R}^{n}
\end{equation}
and provides us with a family of constants of motion of $Z_{H}$. 
This is well known. Let's see what happens if we eliminate a set of the constants of motion from the definition of this structure. In that case, as before we set the associated variables to zero, resulting in the generating function not depending on them anymore. In this way we simply do not have a one-to-one definition of the solution and hence potential solutions depending on various undetermined conjugate variables produce mathematically distinct but indiscernible and fundamentally undetermined trajectories. A complete solution would be transverse to the fibres and hence determine the evolution from one fibre to the next. In a sense this is our gauge connection. However, if the generating function becomes independent of a certain number of variables and their conjugates appear only as arbitrary and undetermined variables the evolution changes dramatically. Instead of a single evolution from fibre to fibre we will have a set of undetermined evolutions, each formed from segments of paths that when combined are continuous but not continuously differentiable. In any case, to make the transition possible, we end up with the same tension between local and global as mentioned above: if the fibre curvature is non-trivial, we will have to take into account the global structure of our bundle and hence the evolution transverse to the fibres that defines our interaction will not be separable. 
Basically this amounts to a not simply invertible map $\Phi$ as described in the above sequence. 
What happens from a physical point of view? 
Even for an apparently classical theory like Maxwell's electromagnetism, it is not possible to separate the path of the light wave once it reached a region of space. There will always have to be an intrinsic correlation between its gauge degrees of freedom that will make the transmission of light possible and that will not have a purely classical counterpart. We can for example imagine a separable sound wave in a fluid medium. We can do that by separating the medium after the wave passed through it. The sound wave will certainly continue propagating. However we cannot possibly separate vacuum after light passed through it because the degrees of freedom of a quantum/gauge vacuum through which the wave passed will be strongly entangled to the degrees of freedom of the emerging wave that correspond to gauge degrees of freedom. To properly separate the vacuum we would have to fight against the local entanglement between regions of vacuum in a quantum field theory, or, in another sense, we would have to fight against the correlation between the unrealised gauge degrees of freedom, which would be extremely energy consuming at best (and probably would result in a black hole). This is also why whenever a gauge interaction propagates, it creates a non-separable structure over the spacetime it connects. In the ER-EPR sense, the spacetime is "sewed" by gauge fields traversing it similar to the way "wormholes" may connect it. 
However, there is yet another interesting property if we think the opposite way. Let us not forget that the principle of special relativity is based on the finiteness and constancy of the speed of light in a vacuum with respect to any reference frame. This principle is true but may have a more universal interpretation once we see that the gauge bundles produce entanglement. In fact, we should define some form of broader principle in which the causal structure should be defined according to the correlations that can be formed by light as passing through space. Such correlations are not classical, not even when a classical Maxwell theory is being used. If gauge connections have a quantum interpretation, then so does the principle of relativity and the definition of a causal structure. These ideas have been analysed only from the perspective of quantum gravity and string theory, but the effects may be extendable to theories we considered far lower on the energy scale, and even for theories we did not see as quantum previously.

\section{fibre bundles and their quantum nature}
As presented in the introduction, fibre bundles have a hidden quantum structure (at least hidden because it hasn't been noticed until now) which may not fully encode quantum mechanics, but which does provide us with some important insights into the emergence of spacetime or of the causal structure that were unknown up to now, and that link gauge theory to quantum information in a new and unexpected way. As quantum mechanics has not yet been fully axiomatised, it is hard to see whether gauge theories encompass all of quantum mechanics or only some of its axioms, but in either case, some of the "quantumness" associated to our understanding of quantum mechanics must be at work in gauge theories, making the emergence of interactions and of the causal structure produced by their propagation across local patches have at least some quantum aspects to them. 
It is important to note that the non-cartesianity of quantum mechanics and by extension of the Hilbert monoidal category is directly connected with the existence of the complex phase and the special rule of calculating the complex probability amplitudes in quantum mechanics, excluding any assumption of a definite intermediate outcome in the absence of a direct detection method for it. The interferences of complex amplitudes in an extended region $R$ in the form described by Feynman through his integral
\begin{equation}
\phi(R)=lim_{\epsilon\rightarrow 0}\int_{R} \Phi(...x_{i}, x_{i+1},...)...dx_{i}dx_{i+1}...
\end{equation}
where $\Phi(...x_{i}, x_{i+1},...)$ is a function of the variables $x_{i}$ defining the path, and where $\epsilon$ defines the time spacing, lead to a construction in which the probability of a path passing through a region is determined by the interference of the complex contributions from all other paths in the considered region. The probability, determined by the absolute square of this contribution will produce correlations that reach outside the causal region of our original path and therefore, by integrating, will provide us with global data. This is how quantum mechanics reaches into the global structure of our manifold.
Let us start with some basics of fibre bundle theory, highlighting the aspects that are regarded by me as fundamentally quantum. First, given the nature of a manifold, we need to be able to associate at each of its points a tangent space with a frame that gives us a basis for it and a connection that gives us the possibility to compare objects (in the special case usually of interest to physicists, vectors) at different points. This brings us to the ideas of covariant derivatives and curvature. This is what the gauge theories have in common and they both can be brought together in the more general construction of fibre bundles. In gauge theories, in general each point on the spacetime manifold $M$ is associated to a complex vector space $V_{x}\cong \mathbb{C}^{n}$ which we identify as an internal space. This structure is one that contains objects linked via linear transformations and hence the underlying operation is that of tensor products and not cartesian products usually associated with sets. Combining therefore two regions in this construction already provides us with a hint towards a quantum component. The matter fields are $V$-valued 0-forms on the spacetime manifold. The basis for each $V_{x}$ is what we know as a gauge. This is analogue to the frame on the tangent spaces of our basis manifold $M$. This gauge is defined locally, on patches of the manifold $U\subseteq M$ and making one such choice is called gauge fixing. Changing the frame is equivalent to changing the reference and is called here gauge transformation. This can be done locally for each $V_{x}$ at each point $x\in U$. All possible changes admitted locally at each point $x$ then are placed together in the gauge group $G$. This is also known as the structure group acting on our vector space $V$ at each point $x\in U$ as 
\begin{equation}
\begin{array}{c}
\gamma^{-1} : U \rightarrow G\\
\rho: G \rightarrow GL(V)\\
\gamma'^{-1}=\rho \gamma^{-1} : U \rightarrow GL(V)\\
\end{array}
\end{equation}
a gauge choice is then associated to a tensor field 
\begin{equation}
(\gamma^{-1})^{\beta}_{\;\;\alpha} : U \rightarrow GL(n, \mathbb{C})
\end{equation}
and the matter fields transform as 
\begin{equation}
\phi'^{\beta}=\gamma^{\beta}_{\;\;\alpha}\phi^{\alpha}
\end{equation}
We can certainly choose the gauge transformations unitarily as all representations of a compact $G$ are similar to a unitary representation and we assume $G$ compact. 
What is important at this point is that we have the linear transformation that implements the change of gauge in one patch and due to this $GL(n, \mathbb{C})$ structure, if we combine two patches we need to consider all possible results of a space of states as being combined via a tensor product, therefore we obtain states on the combined patches that cannot be recovered locally on any of the patches independently. 
If we wish to construct the parallel transport as a way to move on a curve $C$ on the manifold $M$ from point $p$ to point $q$ we construct a map $\xi_{C}:V_{p} \rightarrow V_{q}$. If we choose a gauge, this parallel transport becomes a gauge dependent map $\xi^{\beta}_{\;\;\alpha}: \{C\}\rightarrow GL(n, \mathbb{C})$.
We construct through this the matter field connection linking the tangent space to our manifold at each point and the algebra of linear transformations, again being given traditionally by 
\begin{equation}
\Gamma^{\beta}_{\alpha}(v):T_{x}M \rightarrow gl(n,\mathbb{C})
\end{equation}
The parallel transport is seen as a representation of the gauge group $G$ and the values of the connection are a representation of the associated Lie algebra. The gauge potential is then $\Gamma'=-iqA'$ where $q$ is the gauge coupling. The covariant derivative then is
\begin{equation}
D\phi=d\phi - i q A\wedge \phi
\end{equation}
The connection defines the curvature in the standard way 
\begin{equation}
R=d\Gamma + \Gamma \wedge \Gamma
\end{equation}
from which we directly obtain the field strength 
\begin{equation}
R=-i q F
\end{equation}
However, when we perform the parallel transport we bring together two vector spaces and the parallel transport operator is finally a linear map hence any structure that combines two adjacent patches will not obey cartesian pairing between them. In reality, both patches will be connected in a tensor manner, leading to an overarching structure that will have to include more than just the objects on the two patches. Basically, this simple fact shows that gauge connections basically not only causally connect two adjacent patches but also entangle them in a quantum mechanical sense. 
Let us see this from a fibre bundle point of view. Three components define a fibre bundle: the base space $M$, the bundle space $E$ which also includes a surjective bundle projection $\pi : E \rightarrow M$. The triple $(M, E, \pi)$ is what we call a bundle and $\pi^{-q}$ would be the fibre in a structure usually denoted by $F$. We must have the analogue of an atlas, hence given a collection of open trivialising neighbourhoods $\{U_{i}\}$ covering $M$, having for each a local trivialisation, we have a homomorphism 
\begin{equation}
\phi_{i}:\pi^{-1}(U_{i})\rightarrow U_{i}\times F
\end{equation}
which allows us at each point to work on the fibre only
\begin{equation}
f_{i}:\pi^{-1}(x)\rightarrow F
\end{equation}
This is the local triviality property of a fibre bundle. If we can repeat this for the entire manifold, we obtain a trivial bundle in the global sense and the cartesian property $E\cong M \times F$ holds globally. 
In general a non-trivial bundle is designed to measure global / topological features that are not detectable locally. In gauge theory, as in quantum mechanics, all point-wise defined properties have a linear vector space structure, and hence transitioning from one to the next introduces a tensor product structure that is fundamentally quantum (non-cartesian from the perspective of the spaces of states). Hence, any connection is quantum, given that it cannot fully be described by a cartesian product. Quantum information is usually encoded in the phase of the wavefunction, 
\begin{equation}
\psi(r,t) \rightarrow e^{i\lambda(r,t)}\psi(r,t)
\end{equation}
but also, this phase leads to a fundamental symmetry for the wavefunction in the sense that the results are independent of changes in the over-arching phase, as in fact, by Born rule, one takes the squared norm to obtain the probability. The key term here is "over-arching" in the sense that relative changes in phase between subsystems are visible, albeit not via single local measurements. They do however change the probability distribution of the outcomes and hence connect to global aspects of our manifold. The phase of the wavefunction and the Born rule are essential constructions in the standard interpretation of quantum mechanics and are valid all through quantum field theory and string theory. They also control how the wavefunction "discovers" global information on our fibre bundle and therefore, the fibre bundle construction is particularly informative. 
If we consider two trivialising neighbourhoods on a bundle and their intersection, a fibre over a point in that intersection allows us to construct a homeomorphism 
\begin{equation}
f_{i}\cdot f_{j}^{-1} : F \rightarrow F
\end{equation}
If those homeomorphisms are the left action of an element $g_{ij}(x)\in G$ then $G$ is the structure group of $E$. Hence our gauge group (which is the structure group) allows us to properly connect the neighbouring patches via their intersection. However, in general those intersections contain vector or tensor structures. 
Each $g$ corresponds to a distinct homeomorphism of the fibre. The transition functions $g_{ij}$ bring us from the intersection of the patch to the gauge/structure group
\begin{equation}
g_{ij}:U_{i}\cap U_{j} \rightarrow G
\end{equation}
and hence allow us to define an atlas. The indices $i$ and $j$ denote each patch respectively. Therefore applying the transition function on one of the homeomorphism functions, say $f_{j}$ at the intersection point $p$ moves us from one patch to the next, by means of the relation
\begin{equation}
f_{i}(p)=g_{ij}(f_{j}(p))
\end{equation}
the transition function specifies how the function on the respective patch changes when advancing. 
We can also define a cocycle condition if we consider an intersection point of three patches. $U_{i}\cap U_{j} \cap U_{k}$ which amounts to $g_{ij}\cdot g_{jk} = g_{ik}$ which implies the desired group properties for $g_{ij}$ namely $g_{ii}=e$ and $g_{ij}^{-1}=g_{ji}$. As it seems clear by now, the indices control the advancement on the respective space/manifold while the map brings us from the base manifold $M$ to the group $G$. The action of the group is on the abstract fibre $F$ which is not yet part of $E$ and whose mappings to $E$ depend on local trivialisations. Using these maps one cannot define in general (albeit particular cases are possible) a left action on $E$ because if $G$ becomes non-abelian it will not commute with the transition functions. 
Given the superposed patches that allow us to perform transport on our space, we can ask, at a purely abstract level, what would be the entanglement for such a structure? Of course, in order to ask this question, we need some structure on top of this manifold, one that would be amenable to the classical field space we use, and that would show evidence of non-cartesian behaviour. We do want to keep the added structure classical, in the sense of a classical vector space and the associated classical fields, because the idea is not to quantise a theory, but instead to show that some quantum aspects are required to even define classical gauge fields. It is well known that quantisation can be done in a geometric sense using the three step procedure of pre-quantisation, polarisation, and symplectic form construction. It is also known that to describe a wave-function or a quantum field one can use the concept of a quantum bundle in which the classical vector bundle is tensorialised by an operator algebra related to a bundle of quantum states. Those are constructions that imply the introduction of quantum structures that are supposed to quantise previously classical theories. They also, every time, introduce non-cartesianity in various ways, but in the procedure of quantisation itself. The problem I discuss here is whether the mere construction of a gauge theory that allows a consistently defined parallel transport does already include some form of non-cartesianity that manifests itself in what we used to call a classical field theory. This appears indeed to be the case. In a sense, even our classical world, once it allows for interactions, must be to some extent quantum. We do have two basic principles, one that is the principle of locality, which usually states that global information must be retrieved by some gluing of local information, and the other, which is the gauge principle, that basically tells us how to glue things together. The gauge principle gives us global data in the sense that it gives us enough freedom to locally implement symmetry transformations and to generate connections that allow us to compare objects that are separated. It is the combination of these principles that introduce the concept of gauge interaction and by equivalence, of interaction in general. However, given a theory with classical fields, we cannot simply pair the neighbouring patches on the manifold, as the system described by the field is bound to be paired in a non-cartesian manner. This is why we need to use a tensorial product and we obtain additional gauge freedom. We are basically gluing patches of our manifold by transitioning first through the gauge group $G$, hence generating more freedom in our fields than what a cartesian product would allow. This is not an arbitrary action. We need to do this in order to make any global structure possible. A universe with only the principle of locality but without gauge invariance could probably never be anything except a point singularity. Quantum mechanics relies on the fact that the global state space of a composite system is made up of more than the separated systems themselves, hence it is also forming a global system via tensor products, not cartesian products. Those applied on Hilbert spaces make entanglement possible, but the same type of analysis, in gauge theory, makes any interaction possible. When forming a bundle for a classical field we need to go through a tensor pairing with a gauge group and vastly expand our field space via so called "gauge redundancies". Those extensions are not redundant at all, and indeed we can imagine manifest entanglement in the field-anti-field space or with ghost states. That is in itself interesting, but not truly relevant for this discussion. What is relevant is that the procedure of gluing together patches of a manifold in order to create a global structure and to connect different regions on it, to be able to compare fields or other quantities on them, we require a process of tensoring and expanding the field space that is equivalent to the prescription of pairing in quantum mechanics where cartesian pairing is replaced by tensorial pairing and the space of a composite system is massively expanded. In a sense, the requirement for gauge symmetry, in order to have a meaningful concept of interaction, suggests that a simple cartesian pairing of distinct patches on a manifold would not be sufficient. Of course, it is not to say that the concept of a fibre bundle is just another name for the tensor product pairing on linear spaces. Clearly it is not. But quantum mechanics taught us that the simple cartesian pairing that allowed us to create large structures from elementary components was not sufficient to describe the real universe. We needed a far larger state space, that included entangled states, in order to describe the required phenomena. In the same way, to connect distant objects on a manifold, it is not sufficient to simply pair them, in a cartesian manner, but instead, we need a gauge group and a fibre structure to be able to connect distant objects, and in that way, we have to radically expand the field structure to a series of apparently equivalent fields, separated by gauge transformations. As is the case with the quantum phase, a single gauge transformation is undetectable, but it has been shown that large gauge transformations can reveal additional relevant features that make the distinction between consistent and inconsistent theories and relate them in non-trivial ways [10-11].
To make this more manifest let us look at principal bundles. They are a very common concept in describing gauge theories, having the convenient property of having the gauge group $G$ a topological group that is both an abstract fibre and a structure group. $G$ acts on itself via left translation as a transition function across trivialising neighbourhoods $f_{i}(p)=g_{ij}f_{j}(p)$, the operation $g_{ij}$ itself being a group operation. A fibre over a point is only homeomorphic as a space to $G$ given a trivialising neighbourhood. That eliminates the identity element of our group $G$ which, technically speaking becomes a G-torsor. However, in this situation we can introduce in a natural way a right action of $G$ on the localised fibre
\begin{equation}
\begin{array}{c}
g(p) = f_{i}^{-1}(f_{i}(p)g)\\
\Rightarrow f_{i}(g(p))=f_{i}(p)g\\
\end{array}
\end{equation}
with $p\in \pi^{-1}(U_{i})$. In an intersection of trivialising neighbourhoods $U_{i}\cap U_{j}$ we have 
\begin{equation}
\begin{array}{c}
g(p)=f_{j}^{-1}(f_{j}(p)g)=\\
=f_{i}^{-1}f_{i}f_{j}^{-1}(f_{j}(p)g)=f_{i}^{-1}(g_{ij}f_{j}(p)g)=\\
=f_{i}^{-1}(f_{i}(p)g)=g(p)
\end{array}
\end{equation}
In quantum mechanics we are looking at constituents of larger systems. We can have a wavefunction of a system, composed in the form of a linear combination of wavefunctions of subsystems like in 
\begin{equation}
\psi_{T}(x,t)=\frac{1}{N}\sum_{i}\psi_{i}(x_{i}, t_{i}).
\end{equation}
The phases of each of the subsystem wavefunctions $\psi_{i}(x_{i},t_{i})=e^{i\chi(x_{i},t_{i})}\phi(x_{i},t_{i})$ while unmeasurable if one studies each subsystem separately, do in fact contribute to the overall properties of $\psi_{T}$ and give measurable statistical interferences. This is of course well known, but in a fibre bundle we bring together not subsystems creating a larger system, but instead patches of spacetime (or of our underlying manifold) constructing a path that links two regions in a consistent manner. The same property emerges there, but historically, there was no quantum theory when the first gauge interactions were discovered, hence the research took a different path. However, the situation is very similar. We do not combine subsystems of a system, but in the process of bringing together adjacent patches of the manifold, we bring in a structure that becomes sensitive to the global properties of the manifold, in the same way in which the path integral and the probabilistic interpretation of quantum mechanics makes the relative phases on the component wavefunctions of a system become sensitive to the global properties of the larger composite system. This is described in gauge theory by a fibre bundle, and it comes as no surprise that the wavefunction or quantum field approach to quantum mechanics is also amenable to a fibre bundle interpretation. We just seem to be more accustomed to the wavefunction/field viewpoint of quantum mechanics. 
\begin{widetext}
\begin{equation}
\begin{array}{ccc}
\begin {tikzcd}[column sep=small]
	& G \arrow[dr] & \\
 \bigcap_{i}U_{i} \arrow[ur] &			     & (E,M,\pi)
\end{tikzcd} &
\Longrightarrow
&
\begin {tikzcd}[column sep=tiny]
	& e^{i \chi(x_{i},t_{i})} \arrow[dr] & \\
 \sum_{i}\psi_{i}(x_{i},t_{i}) \arrow[ur] & & \psi_{T}(x,t)
\end{tikzcd}
\end{array}
\end{equation}
\end{widetext}
In both cases we expand the state space in order to be able to connect local to global information.  It wouldn't have been otherwise possible to move from one patch to the next in a consistent manner. 
As said previously, quantum mechanics relies on a strong axiomatic support materialised by the non-cartesianity of the pairing operation between subsystems, and by extension on the interference of complex amplitudes of unrealised intermediate states. This offers it a unique insight into global properties of systems and manifolds by means of a probabilistic approach. The other way in which quantum mechanics is different from the classical way of thinking is the extension of observables to operators that encode all possible outcomes of an observation, at a fundamental level. Indeed, it is this property that gives quantum mechanics its probabilistic nature and allows us to implement Born rule to derive measurable results. In fact, this property is intrinsically related to the first one, namely the existence of the complex phases, and allows quantum mechanics to infer non-local information via quantum correlation. General relativity doesn't seem to have such a probabilistic approach. However, gravitational observables are non-local. Indeed, it has been shown that in general relativity and finally in quantum gravity as well, we do not have local gauge invariant (diffeo-invariant) observables. One could imagine that this aspect is the analogous situation of the spreading out of the quantum observables in basic quantum mechanics. In the same way in which we expand classical observables that have only one possible outcome (the classically "real", absolute property of the system) to quantum observables with their operatorial underpinning and no absolute outcome for unmeasured observables, in gravity we need to spread out the observables, not over the space of various unrealised possibilities, but over the spacetime itself. 

In the following image (Figure 1), the vertical plane plays the role of a quantum region where the dots on the circle represent the possible eigenvalues of our quantum observables. The intersection line of the two planes is the classical realm, where only one outcome is the "true" (absolute) one for each observable, and spacetime gauge invariant observables should be point-like (hence we are in a flat Minkowski spacetime). The horizontal plane would then be the spacetime, where gravitational gauge invariant observables are non-local and hence rely on spread-out observations at the dots on the horizontal circle. Indeed, if we ignore the dots on the vertical plane we miss all the interferences of quantum mechanics and it becomes impossible to understand the probabilities we would obtain, leading to the well known paradoxes of classical mechanics. On the other side if we look at the horizontal plane, if we ignore the dots on the horizontal circle we are not capable of detecting gravitational observables and hence are insensitive to curvature, making us ignorant with respect to the global structure of spacetime. With this analogy in mind, could we imagine that the curvature in the horizontal plane is related to the entangling structure of quantum probabilities in the vertical plane? 
\begin{figure}
  \includegraphics[width=\linewidth]{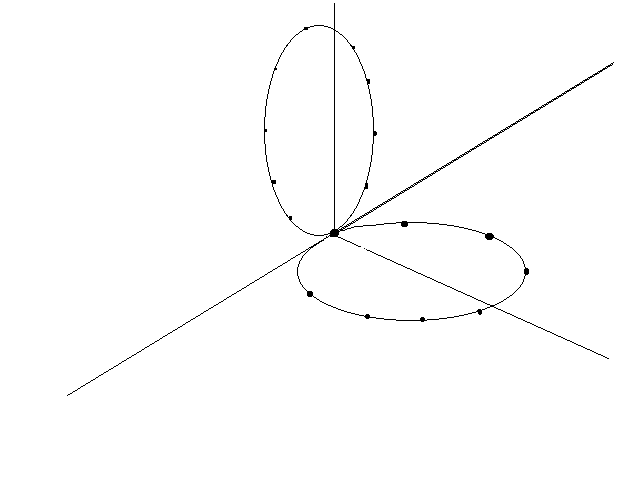}
  \caption{Imagine two observables, in the vertical plane a quantum observable, in the horizontal plane a gravitational observable, and at the intersection of the two planes, the classical line}
  \label{fig:gauge1}
\end{figure}
Figure \ref{fig:gauge1} shows the idea presented above. 

In fact, the two should describe the same global structure from two different perspectives, one quantum statistical (harmonic analytic), the other one geometrical, based on curvature and gravitational observables.
This would link two relatively different branches of mathematics. Let us remember the Atiyah-Singer index theorem. There, the analytical index of a differential operator is linked to the topological index of the underlying compact manifold. This theorem linking differential operators to topological structure is a broad generalisation of a series of other theorems including the Chern-Gauss-Bonnet theorem which links topological invariants to polynomials in the curvature form. 
If the two methods of detecting global information are to be equivalent, we expect another connection, between the eigenvalues and eigenfunctions of our observables (harmonic analysis) and the spacetime curvature (or some functions of those, hence geometry). Indeed, one could imagine a density matrix approach in which the pure and mixed states, entangled or not, give us enough structure to connect them with integrals over paths in spacetime that lead to curvature forms. A study of such a phenomenon would seem to be quite interesting. Density matrices are already a nice unification of thermodynamic probabilities and quantum probabilities. I call thermodynamic probabilities, those probabilities required to describe the likelihood of different outcomes emerging from our ignorance of some underlying absolute results. I call quantum probabilities, those probabilities required to describe the likelihood of different outcomes emerging from the lack of any absolute intermediate outcomes in the processes analysed. Of course the term "thermodynamic" has a broader meaning, and of course one can imagine quantum thermodynamic properties when those broader meanings are included. I am probably not imaginative enough with new terminology here, and I apologise for that. 


\section{Where is the curvature?}
Analysing figure 1 we can ask ourselves what is the connection between a purely quantum and a purely gravitational description. It seems manifest that classical physics is a limiting case of both situations, a case in which no access to global information is possible and every observable has only one possible outcome at any time and in any experimental context. Clearly that is too limitative for both general relativity and quantum mechanics. Would there exist some connection between a purely gravitational and a purely quantum description? In a sense, I can imagine the relation between the two to be one of duality, in some sense in the way in which the tangent and the cotangent bundles are isomorphic and relate to some underlying manifold. This is of course only an analogy, but the general idea is that gauge theories, with their bundle structure, are to some extend dual (and maybe isomorphic) to a quantum description. Attempts to quantise gravity would therefore fail because gravity and general relativity would be a low energy limit of a quantum theory of gravity that would still contain quantum elements that were not recognised as such up to now. Gauge field theories also have apparent quantum elements to them, which created the gauge fixing problems in quantising gauge field theories, problems that are not yet completely solved if we think at Gribov ambiguities and the confinement problem. The mathematical tools used to describe the two can be harmonised in the sense that both rely on very peculiar interpretations of what we mean by "physical". In both quantum and gauge theories we have certain variables that appear to play a role in analysing global results but are not explicitly realised "physically". Another key element is discernibility. In gauge theory we say that the different although equivalent gauge paths are brought together in an equivalence class, the classification of which is done in terms of BRST cohomology. In any case, while mathematically distinct, we talk about gauge equivalent observables as identical from a "physical" point of view. We also talk about gauge variant observables, namely those variables that depend on the arbitrary choices of gauge "frames", as "non-physical" in the sense of being ambiguous to some extent. A similar discussion we seem to have in quantum mechanics. When integrating over intermediate configurations, we integrate them in order to form a probability amplitude by realising that, although they do appear this time to be "physically different", the choice of the experimental setup is unable to discern them. We do indeed in this case integrate over what we may call (from an external point of view) "distinct" positions or momenta, but from the perspective of the experimental setup or the context choice we made when we designed the experiment, those "physically different" positions and momenta are quite equivalent, as there is no way in which a "which path?" question could be answered without fundamentally changing the experiment.
We basically created, by building a specific experimental setup, a situation in which some additional "gauge freedom" is implied and this time so called "physical" configurations that may look distinct, are simply impossible to discern. My point of view is that the distinction between what means "physical" and "non-physical" here is just semantics, and from the perspective of Nature, both quantum and gauge theory have their own unrealised, indiscernible configurations which cannot be ignored if global information is to be extracted from the system. It seems like Nature gives us more degrees of freedom than what we would imagine is necessary to discuss about the dynamics, simply because various context or frame choices are allowed and global information plays a role even in local experiments. 
I discussed in ref. [17] a possible interpretation of the curvature in a dual quantum representation. We should remember that all gauge interactions are based on field strength tensors $F_{\mu\nu}$ which play the role of a curvature in the inner space. Without such a curvature on the gauge connections we wouldn't have most of the elementary interactions we see. The only alternative would be the Chern-Simons terms relying on a flat connection with topological structure. However, to probe such a curvature directly we would require either a path in the form of a closed loop, or two distinct paths intersecting in some points. The impact of this observation I tried to analyse in an old article of mine [25] and then [26], but my new understanding on the gauge - quantum relation may shed new lights on that and also show how curvature in inner space may play some interesting roles in solving computational problems considered hard.

If we try to understand the gauge-quantum duality from the perspective of the path integral approach, we have to think in terms of what type of paths appear in the path integral quantisation prescription and what is their relation to the integration over equivalent paths in gauge theory. The Feynman path integral involves contributions from paths that may even be nowhere differentiable, as all possible paths do contribute to the integral. Therefore, it is not the well behaved paths that are typical in the Feynman path integral approach. In this approach, the time slicing parameter, let us call it $\epsilon$ plays the role of a regulator for such an integration. We know of attempts to reconstruct the Feynman path integral approach in which only well behaved (differentiable) paths were considered. In that case, as done in ref. [29] the high energy, small time-scale behaviour is rather different from the behaviour of normal quantum mechanics. Even with such a construction, the transition between paths can be obtained by means of transformations that may, in the "worst" (but probably more realistic) case scenario be also depending on arbitrary choices of the experimental setup in each position. Now, the experimental setup results in a certain arbitrariness in choosing the starting and ending point of each segment of our path. This is what ultimately leads to the derivation of commutation relations in quantum mechanics. 
The question rises: does a similar situation occur in the case of gauge invariance? We have seen up to now that gauge invariance leads to the same type of indeterminacy as in quantum mechanics, when analysing the Hamilton-Jacobi equation and the associated incomplete integral. In fact it has been shown that the evolution equation in the case of a gauge transformation also contains an arbitrary function and each such choice in each point for such a function contains a series of possible constructions that are indeed nowhere differentiable. It is probably important to notice that while the gauge transformations link the paths locally, each gauge transformation, being fundamentally local, can be chosen to be different in the next step, mimicking exactly the same type of freedom we have in the quantum counterpart. The ability to perform transformations between paths in the gauge case is an expression for the freedom of choosing different gauge frames as allowed by the gauge symmetry of the theory. The ability to perform transformations between paths in the Feynman path integral case is an expression for the freedom of choosing any possible path allowed by the experimental setup. Ultimately it could be argued that the two freedoms are equivalent but it is indeed interesting to see how such a proof could be performed in a mathematically rigorous sense. In both situations however, physically, the paths are equivalent. In the gauge case due to the fact that the theory is independent of arbitrary choices of frames, while in the quantum case, because the experimental setup is constructed such that the "agents" (electrons, photons, etc.) could choose any of those paths, without being able to explicitly "decide" whether it was one path chosen or another.

\section{an infinite vs. a finite number of representatives in a class}
The main problem when quantising gauge theories is that, if no gauge fixing is implemented, the integral obviously diverges. Not much thinking has to be involved into noticing this: we are literally integrating over an infinite number of equivalent configurations. Only that those configurations are probably not as "equivalent" as one may wish. Given the quantum nature of those gauge configurations, there may be some structure to which those "gauge choices" could make us attentive. When integrating over a gauge fixed problem, we do recover the global BRST symmetry.
What happens is that we basically "solved" the problem of over-counting the degrees of freedom by "infinity" by restricting the choices of each equivalence class to one single representative of each class. Integration works well in this case, but there is still a lot of structure involved that we may as well miss. For example, we cannot control an infinite number of elements integrated over in each equivalence class but what about a finite number of them, but not only one? 
I was thinking about this problem since ref. [25] and [26] where indeed, I tried to control the integration over several paths in which two (or more) choices of gauge representatives were made simultaneously. While it looked strange, it appeared that at least one NP-type problem was solved quite quickly (at least at the level of a simulation, there was no way of accessing a quantum computer of any sort at that time and at that time, I actually didn't know it was a quantum problem to begin with). What I did there was to pick more than one but less than an infinity of representatives in a gauge "equivalence" class, and exploit them in order to create various types of symmetries between those emerging paths. I think it should be best to refer to the reader to the figures 2 and 3 added here. 
\begin{figure}
  \includegraphics[width=\linewidth]{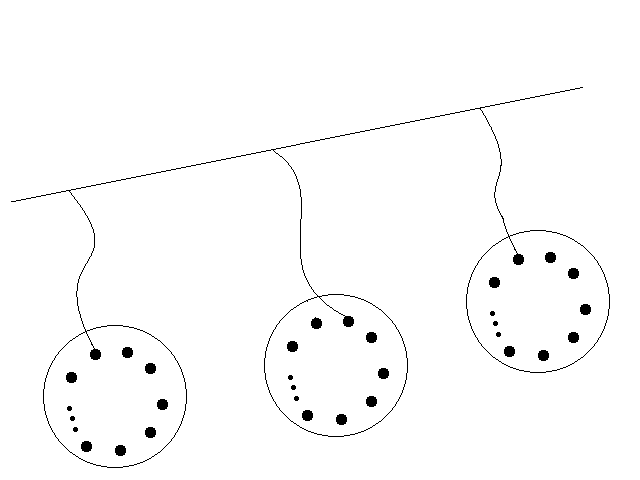}
  \caption{Usually, gauge fixing procedures work like this: a representative of each equivalence class is taken at each point, and the path integral is performed only over gauge inequivalent states. This makes the integral finite and well normalised. However, by doing so one loses information that can emerge only from understanding correlations between different elements of the class.}
  \label{fig:gauge2}
\end{figure}

\begin{figure}
  \includegraphics[width=\linewidth]{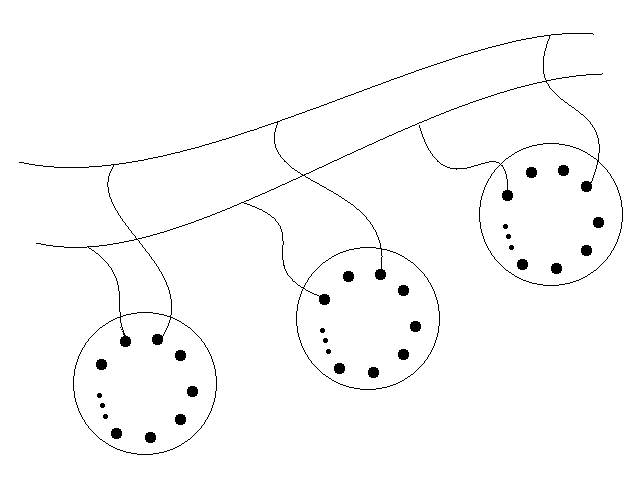}
  \caption{Instead of reducing an infinite number of "equivalent" gauge configurations to only one, we could reduce an infinite number of gauge configurations to, let's say, two, or any finite number. This is possible and normalisable. If any class contains an infinite number of choices, if we can pick one representative, we could as well pick two. In principle we should be able, in this way, to generate a broad spectrum of symmetries between the paths, facilitating the solution of some complex problems. Those amount to, as I understand now, different experimental setup configurations of the "quantum machine" that implements the experiment. A situation in which a path crosses the same gauge class twice is in the Gribov problem, both intersections having to be taken into account to better understand the confinement problem.}
  \label{fig:gauge1}
\end{figure}
The mathematics of this has been described in ref. [25] and [26] in great detail so I will just comment here on the interpretation thereof and of several things I am now understanding slightly better compared to [25] and [26]. First, in ref. [25] and [26] I wasn't aware of the connection I see now between quantum mechanics and gauge theory, therefore I considered the simultaneous double choice of representatives a mathematical tool to create additional structure in the problem. With only one path going through each equivalence class once, we are somewhat limited in terms of symmetries we could add to the problem. With an infinite number of paths, resulting from the "spurious" integration over all gauge configurations we are certainly overwhelmed by the additional structures and symmetries we recover, with no chance of figuring out what is relevant and what not. But with a finite number of such configurations, a number of controllable features can be added to the setup making the calculations in many aspects simpler. One argument raised regarding article [26] was that finally I do not obtain the same theory, which at that point I argued against, as the only change I introduced was (given a proper normalisation and according to the point of view at that time) a "spurious gauge". Of course, now I understand that even with proper normalisation of the problem, the situation is not quite the same, as two gauge trajectories do have the possibility of encoding different potential outcomes described by unrealised internal states. Therefore a one path through each class configuration will contain less information than a two paths through each class configuration, even though the inner states themselves are not observable and (given an experimental setup) equivalent. Moreover, two paths in the same equivalence class can probe correlations within the class that were not clear before. We should be able to probe the structure of the equivalence classes themselves, as the elements therein are only equivalent with respect to the rules we prescribed them to be equivalent to, but they may possess additional properties that distinguish them and make them show interesting structure. Topological features of those classes create Gribov copies and other features, but it is not excluded that other features, like superposition, entanglement, braiding, or even intersection may have non-trivial physical implications. Also, it was my opinion in ref. [25] that a gauge modification of this type could imply having an efficient algorithm for an NP problem. This is as for now not precisely known, but it is believed to be possible that certain classes of problems not solvable on classical computers can in principle be solved on quantum computers [27], [28]. 
In the case of Gribov copies, the non-linearity of the gauge trajectory leads to a reversal and the intersection of a class several times. Therefore it clearly is possible, and unavoidable in the context of QCD, that certain classes are probed twice via different but gauge equivalent configurations. The result is a construction that better takes into account the problem of confinement. This seems to be however a topological feature, but there could as well be geometrical features of the classes that would become more manifest if a finite but not one paths probing each class were allowed. Up to this point I can imagine a better approximation to the confinement potential obtained in this way, although a comparison with a lattice model would be desirable. More about this will be discussed in a dedicated article.

\section{some philosophical thoughts}
Because in this article I am connecting classical gauge theories on one side with quantum field theories on the other, it becomes interesting to comment what are the implications of such a connection. On the other side, one may ask how such a connection can be relevant to some interpretations of quantum mechanics, like for example the Bohmian or pilot wave interpretations. We have to go back to the goals of Bohm while trying to understand quantum mechanics from the point of view of hidden local or global degrees of freedom. In that context the general assumption was that the non-determinacy in quantum mechanics originates from the lack of knowledge of some underlying degrees of freedom that completely determine all properties of the system, and that those degrees of freedom are actually physically realised. This point of view resembles that of classical thermodynamics, in which the only reason for fluctuations is an empirical lack of knowledge of the actual outcomes of some observables. These observables may as well be non-local, but their main property was that they were physically realised, albeit unknown to us, leading to a fundamentally full determination of quantum mechanics. In this sense, quantum mechanics would be transformed into a peculiar classical theory, where some non-locality was present but all observables that we can imagine about a system would have one and only one outcome the moment they are realised physically. Instead, historically, quantum mechanics remained a unique theory in which not all intermediate states have a clear unique outcome that is actually realised, transforming quantum mechanics into a theory in which potentialities are being considered as having a direct impact on the results of the theory, despite them not being actually realised. 
My approach in this article is that Quantum mechanics retains its fundamental indeterminacy, and that in reality such intermediate states are indeed not realised. From this point of view my approach is closely related to the so called Copenhagen interpretation. However, there is a parallel that seems to develop between quantum mechanics (or quantum field theory) and classical gauge systems. Does that mean that a re-interpretation of quantum mechanics indeterminacies as gauge indeterminacies makes the theory classical? I strongly believe not. My view, as presented in this article, is that indeed, the gauge nature of interactions basically implies some hidden (unnoticed up to now) quantum aspects that are associated even with what we used to consider classical gauge theories and classical physics. In the same way in which the choice of a frame or of a gauge determines some properties, while making others less determined, the choice of the context of a quantum experiment makes some observables have fully determined outcomes, while others can only fluctuate around various possible outcomes. 
In that sense, as a dynamical theory, quantum mechanics has to carry with it a series of possible outcomes, like, for example in the Feynman approach, an integration over all possible inner states, because all will contribute to the final amplitude. As I showed in this article, this is a feature shared with gauge theory, where, also, we have to carry in the dynamical equations solutions that depend on arbitrary choices of frames. Both gauge and quantum theories carry with them more information than what would be classically considered: in quantum mechanics, potential realisations of the inner states, and in gauge theory, potential choices of frames (or gauges). The element of choice is relevant in both cases: although it seems natural that we can always choose a specific gauge, some aspects may become invisible if the choice of gauge doesn't manifestly express them. In the same way, in quantum mechanics, we make a choice of the observable context, therefore specifying our perfectly determined set of observables, leaving the rest undetermined, but not irrelevant to the calculation. 
One characteristic of Bohmian mechanics is that a so called Bohmian or quantum potential appears to guide the particles. This potential is non-local and emerges from the real part of the Schrodinger equation written in polar form. Such a potential has historically been shown to have some very abnormal behaviours. Authors have claimed it corresponds to backwards-in-time information transmission, aberrant non-localities, and even time travel. In any case, if such an object is considered physically realised, it is of course absurd to demand it to have such properties. If however, we consider it as the result of gauge choices that have to propagate and to encode global gauge properties, like in the case of arbitrary functions encoding choices of frames, we do not have such problems anymore. It is not impossible to discuss in terms of a generally covariant description involving gauge variables in which time evolution is reversed. For example in an inner loop of a Feynman diagram, we can easily (and with no harm) consider particles moving directly in time while anti-particles moving backwards (this is just one possible example). This of course implies no physical transmission of information backwards in time, just a simple choice of a convention and the follow-up of its consequences in a global context. In a sense, such assumptions link a perturbative approach related to the series expansion in terms of Feynman diagrams with certain global properties, which, interestingly enough, occur in what we call "quantum corrections", that are described by means of inner diagrammatic loops. 
\par The impact of global gauge properties on perturbative calculations and the quantum dual interpretation has an important role in understanding for example QCD confinement, a subject of a future article of mine. Re-writing the Hamilton Jacobi equation using this method leads indeed to an additional term 
\begin{equation}
\frac{\partial S}{\partial t}=-[\frac{|\nabla S|^{2}}{2m}+V+Q]
\end{equation}
where 
\begin{equation}
Q=-\frac{\hbar^{2}}{2m}\frac{\nabla^{2}R}{R}
\end{equation}
where again $R$ is the absolute value of the wavefunction, while $\frac{S}{\hbar}$ is its phase
\begin{equation}
\psi=R\;exp(iS/\hbar)
\end{equation}
This potential can indeed be found from gauge arguments alone, and in fact can be associated with a curvature term that comes only from the part that can be arbitrarily deformed in the usual gauge potential. This allows us to construct a quantum theory in which the outcomes remain undetermined, yet, due to their impact on the gauge space they do have an impact on the amplitudes coming from global (or non-local) effects. 
These aspects have been treated in detail in ref. [17], by the same author, and is currently in an advanced stage of development. It is in ref. [17] where I also discussed the Feynman integral type construction obtained by reducing the integrals of motion, resulting in a series of undetermined paths which we have to consider as intermediate unobservable states of the system. We can then use a general covariant construction over those paths involving transformations of the canonical coordinates by means of connections. This happens in the extended gauge space and we can define an associated parallel transport in this inner space.

\par If we look at the Feynman path integral formalism and at the possibility of expanding the action by means of gauge terms, we see further similarities between the two ideas. In the path integral approach, we also obtain a series of potential paths, which are fundamentally indiscernible while each of them having a different numerical value for the action functional. In a sense, those paths are receiving a weight given by the action functional over the respective path, but are being introduced in the quantum formalism as a phase and hence, by the same quantum formalism, they remain indiscernible and unobservable (indeterminate). It is worth mentioning that they are physically indistinguishable although they do differ in terms of the weights assigned to them. The same experimental setup can produce any of the possible outcomes appearing as eigenvalues of an observable. After they are being realised, of course they can be distinguished from others through their values themselves and the probability of occurrence, but before the experiment is performed, given the same experimental setup, any of the eigenvalues could occur, and hence the experiment couldn't distinguish them all by itself. The same is valid in the case of the gauge choice. A given gauge can mathematically be distinguished from another, but in terms of physical realisation (say, by means of an equation of motion) one gauge choice is equivalent to another, that is, until large gauge transformations are employed and global effects make it clear that physical distinctions among gauges do exist. The solution to this, as performed in the Feynman formalism, is to construct the amplitude by integrating over all such possible and equivalent (indiscernible) paths, and obtaining the probability of transition from the initial to the final point (or state) by applying the usual Born rule. The same situation occurs in gauge theory. Indeed, gauge variant observables received a rather undeserved bad reputation as being "unphysical" although they clearly contribute to a plethora of physical effects. For a discussion about this subject, see ref. [17]. Briefly, let us think about what makes an observable gauge variant, and what does it mean? A gauge variant observable basically depends on the choice of a gauge, or a choice of an arbitrary function. In special relativity we would call this a choice of a reference frame. This would make the concept of length a "gauge variant" observable, or, as we say in relativity, a reference frame dependent observable. Sure, length has the tendency of contracting according to the speed of the reference frame considered for the observation, but that doesn't make the concept of "length" unphysical, it only makes it ambiguous once one wishes to discuss the problem independent of reference frames. We continue to have spacial length, even in special relativity, although we know of the Lorentz invariant interval defined in terms of spacetime. A very similar situation occurs in gauge theory. In fact, for a somewhat customary reason we are used to call gauge variant observables "unphysical" when instead we should just call them "ambiguous". The use of the term "unphysical" became so common that gauge variant quantities are automatically deemed to be even "unreal" or "unnatural". In any case, there are a series of gauge variant quantities that, while being ambiguous, cannot be excluded from physics. One of them is quark colour. We call quark colour unobservable because it is gauge variant, and indeed it is. The colour of quarks changes as we apply the SU(3) type transformation of gauge, meaning that observers making different choices of gauge transformations will detect different colours. But this is really all there is to it: given an arbitrary choice which a group of observers decide to share, the measurement of the colour of a given quark will stay the same. In that sense, the colour property is not fully determined, in the same way in which a path in the Feynman approach or in the construction I presented in this article is not fully determined unless one adds some arbitrary choice to it. That doesn't mean we have to reject the whole property and in fact in QCD we do not, otherwise it would be hard to justify a theory called quantum "chromo"-dynamics. While I do know that these facts are intuitively understood by experts in the field, the "stigma" that something gauge variant is "non-physical" remains.
A collective pre-established choice or a pre-determined strategy on which observers may agree upon in the beginning can bring additional determinacy into the problem. I would like to make the reader aware of ref. [24] where the initial choice of a specific strategy is capable to fully determine the outcomes that would otherwise be inaccessible via a simple signalling method. 
 As is the case in quantum mechanics, a choice of gauge (or a choice of arbitrary function) allows us to complete the experimental setup in the same way in which the choice of an axis in the spin-0 decay experiment allows us to define the projection of a spin and subsequently determine the specific orientation in a specific outcome. The main point is that a choice of context (be it a reference frame, gauge, or a specific experimental context of a quantum experiment) does add the required information that was not initially and naturally attached to the system, and after such a choice is made, the information about specific properties can fully be determined. In this sense, we see that special relativity (reference choice), gauge theory and interactions (gauge choice) and quantum mechanics (choice of a context of observables) all depend on getting a specific context in order for certain properties to be fully determined. We can think of special relativity as allowing for arbitrary functions in terms of choices of reference frames, while presenting us with an "absolute" Lorentz invariant spacetime interval (or proper time), we can think of the gauge theory of interactions as allowing for arbitrary functions in the choice of gauges, while presenting us with "absolute" gauge invariant observables, and we can think of quantum mechanics, allowing for arbitrary functions in terms of the phases of wavefunctions and the indetermination of certain observables, while allowing for a choice of observables that are fully determined given a context. In a sense, quantum mechanics surpasses the previous two theories in the sense that it basically says that no matter what choice of perfectly determined (commuting) set of observables one makes, one will always have another set for which only imprecise information about their outcomes will exist. Special relativity takes away a set of absolutes (say spacial distance and time intervals) but gives us another absolute (the spacetime interval, the proper time). The usual interpretation of gauge interactions also takes away a certain type of absolute, associated to generally defined observables, while giving us a certain restriction, of gauge invariant observables, long thought to be the "absolute", unambiguous, "real" observables. Quantum mechanics is in a sense closer to reality, as it takes away the absolute idea that every observable must have a single outcome, but doesn't set up any absolute to replace this with. Any choice of absolutely determined (commuting) observables will have as a counterpart other observables that will remain undetermined. This approach incorporates gauge theory, and potentially also relativity. Given the fact that quantum mechanics seems more natural from our current understanding, one may ask what impact such a powerful relativisation would have if we trace back our steps and use the same analogy to go to special relativity? In fact, if one tries to see a quantum structure in the usual classical gauge connections, as I showed here that it is the case, we may obtain a further "relativisation" of the absolutes of special relativity, given the fact that the light beams from the "light beams and clocks grid" of special relativity are basically gauge connections which are ultimately quantum in nature and produce the entanglement of adjacent regions of spacetime. 
 Another question one may ask would be what would be the effect of understanding the connection between gauge and quantum on the existence of the gauge groups we know in the standard model. 
 There are as of now, no clear reasons as to why the groups of the standard model should indeed be those we have discovered in the 20th century. There are very interesting geometric and topological approaches in string theory, but they all suffer from a very similar problem: there are too many possible outcomes, and very few constraints that would unambiguously restrict our choices to the standard model gauge group. It is indeed not hard to derive the existence of low energy gauge groups in string theory, and in fact it is also possible to come up with something very close to the standard model gauge group structure, but it is not at all clear why a specific gauge group is the one and only gauge group appearing at low energies. We correctly assumed that a series of symmetry breaking processes, culminating with the spontaneous (hence at the level of the solutions of the equations of motion, and not at the level of the action) gauge symmetry breaking a la Higgs describe the standard model physics properly. While this idea definitely needs more exploration, it would be interesting to see whether, if the thought process leading us to this point, followed backwards, would lead us to a better understanding of a potential quantum origin of the standard model gauge structure and of the Higgs symmetry breaking mechanisms. Other symmetry breaking mechanisms can be led back to entanglement. For example, the explicit, gravitational instanton triggered symmetry breaking, contributing to the mass of axions, has been traced back via ER-EPR to some quantum effects [22]. A rather different mechanism also originating in quantum mechanics seems to be associated with the spontaneous symmetry breaking, and at the same time with the problem as of why the gauge structure of the standard model is the way it is. 
 Simply explained, the cosmological evolution might have naturally added a specific context choice that further determined this and only this gauge group as the standard model group due to some results that may derivable from quantum computation theory. That may amount to entanglement between widely separated regions of the universe. 


In the process of defining a connection and a curvature of the underlying space constructed using the unobservable variables, we may follow ref. [18] and demand that the parallel transport of the length vector in this space changes by a linear law. The resulting geometry is a Weyl geometry and treating this connection as a gauge field we obtain a modification that leads to the additional quantum potential. As opposed to ref. [18] however, in my approach, this is strictly the result of unmeasurable gauge terms and the curvature of the space associated with them. As opposed to [18] where this potential appears as a result of real spacetime fluctuations (and in my opinion, such a result cannot account for the quantum effects we measure), in this case, the quantum potential appears as a result of gauge arbitrariness and the possibility of defining a space and a curvature over unrealised inner states. As said, more details on this aspect are discussed in [17]. 
In terms of applications, this approach can have some interesting results, which I will only briefly mention here, as they are essentially the subject of future articles. 
First, if we can discuss gauge properties in terms of quantum properties, according to this gauge-quantum duality, it is interesting to ask what is the quantum equivalent of some of the well known dualities in field and string theory. Such a fundamentally non-local construction emerging from gauge choice arbitrariness could be translated in terms of entanglement and lead to a quantum information interpretation, for example for T-duality. I will discuss this extensively in ref [19].
As mentioned in ref. [19], ghost fields are usually used in functional approaches to non-perturbative effects in quantum field theories (and in particular in quantum chromodynamics) to consistently transport gauge choices across the gauge space. In all cases, their effect is considered spurious, but there is no known way of performing calculations without them. Their presence, for example in gluon propagators or in vertices is an indicator of these constructions being off-shell and even potentially confined. However, a far more important application is given by the so called Veneziano ghost. This is the subject of a completely new paper altogether, but I think a glimpse to it could be given here as well. Non-abelian gauge theories present us with a fundamental ambiguity due to the existence of multiple solutions to the gauge fixing condition. These ambiguities are called "Gribov copies". In fact, it has been shown by [20] that indeed this ambiguity is the result of the very nature of non-Abelian theories, and no unambiguous gauge fixing is possible due to the global properties of such theories. Gribov's proposition was to restrict the range of functional integration to avoid multiple copies. This of course is not a globally sensible approach. However, such a restriction does indeed change the type of interaction making it confining. According to [21] the Gribov copies naturally arise due to the compactness of the gauge group. The way we construct the vacuum of the theory must therefore accommodate their existence. [21] proposes therefore to instead sum over all Gribov copies as well, which also leads to a confinement of gluons at large distances. In fact, the existence of Gribov copies and the emerging discontinuities of gauge potentials are required to accommodate for the vacuum configurations with arbitrary Chern-Pontryagin index. This index in even dimensional spacetime is associated with topology changing transitions. Usually topology in gauge theory is only discussed in terms of semi-classical solutions (instantons, monopoles, etc.) ignoring the non-perturbative QCD effects. When performing perturbative expansions, we do need to fix the gauge and the existence of Gribov ambiguities makes it therefore impossible to isolate perturbative theory from the topological nature of QCD [20]. In gauge theories, topology is important due to the compact nature of the gauge group. The relevance of topology in this context therefore brings us to the connection between topology and confinement. Compact gauge groups can be homotopically mapped into the spacetime manifold. The map from $SU(2)$ subgroup of the gauge group to the Euclidean spacetime sphere $S^{3}$ describes a classical instanton solution. If we transform to Minkowski spacetime, instanton solutions correspond to tunnelling connections between degenerate vacua with different Chern Simons numbers
\begin{equation}
X(t)=\int d^{3}x K_{0}(x,t)
\end{equation}
with $K_{0}$ the temporal component of a topological current 
\begin{equation}
K_{\mu}=\frac{1}{16\pi^{2}}\epsilon_{\mu\nu\rho\sigma}A^{\nu\;a}(\partial^{\rho}A^{\sigma\;a}+\frac{g}{3}C^{abc}A_{b}^{\rho}A_{c}^{\sigma})
\end{equation}
This topological current is not gauge invariant, but its total derivative is
\begin{equation}
Q(x)=\partial_{\mu}K^{\mu}
\end{equation}
A non-vanishing correlator $\Bracket{Q\; Q}$ at zero momentum, i.e. a non-vanishing topological susceptibility, only appears possible if there is a massless pole in the $\Bracket{K\;K}$ correlator. Such a pseudovector pole is associated to the so called Veneziano ghost. More precisely, Veneziano proposed 
\begin{equation}
\mathcal{K}_{\mu\nu}(q)=i\int d^{4}xe^{iqx}\Bracket{K_{\mu}(x)K_{\nu}(0)}\xrightarrow{q^{2}\sim 0}-\frac{\chi^{4}}{q^{2}}g_{\mu\nu}
\end{equation}
It has then been seen that such a correlator appears as the effective interaction between the gluon and the Veneziano ghost. Solving the Dyson-Schwinger equation using solely this coupling leads to a dynamically corrected gluon propagator called by the authors "glost" given by 
\begin{equation}
G_{\mu\nu}=\frac{p^{2}}{p^{4}+\chi^{4}}g_{\mu\nu}
\end{equation}
An important aspect is that the topological $K_{\mu}$ current and its correlation function are gauge dependent as such. They are also topological in nature, and are, according to my gauge-quantum duality, describable by means of quantum information tools, being most likely based on entanglement. It is not terribly surprising, given this interpretation, that the topological current has in the context of abelian field theories, the interpretation of magnetic helicity, and corresponds (as a Chern Simons term) to the knottedness of the magnetic flux lines. In non-abelian theories, the quantum entanglement nature should be even more manifest, and a connection between quantum information and topological properties of non-abelian gauge theories can be more explicitly derived [22]. In any case, these and more applications should be the subject of future research. 
All in all, it is my view that in order to properly approach non-perturbative effects, we have to better understand this new connection between gauge field theories interpreted quite generally and from first principles, and the principles of quantum mechanics, maybe even related to a proper interpretation of quantum mechanics at a foundational level. Therefore, quantum interpretations do not appear to be simply "ideological" or "metaphysical" discussions, as their interpretation may have a direct impact on general calculations in non-perturbative physics. In fact, quantum entanglement seems to behave not only as a quantum informational tool, but also as a mathematical tool capable of providing us access to global information that is otherwise inaccessible. The so called quantum potential, introduced by Bohm in a way that may not be fully correct from our modern perspective, may however have an important role in understanding some of the most important topological features of QCD due to its re-interpretation as a gauge property of the system. This article makes a step towards this by presenting a duality between quantum mechanics (or quantum field theory) and (classical) gauge theory.

One of the most interesting questions asked since the beginning of quantum mechanics was how can we obtain classical physics as an extreme limit of quantum mechanics. There are of course several well known limits, for example the limit in which the Planck constant is small $\hbar\rightarrow 0$. Another limit is related to the inner space gauge curvature as presented in ref. [17]. This is discussed in more detail there. However, if we think about the ability of quantum phenomena to be sensitive to global structure and consider this as being the main common feature characterising both quantum mechanics and general relativity (and by extension other classical gauge interaction theories) then we have to consider the "classical" limit in which each observable has one and only one possible and fundamentally fully determined outcome as an approximation, or a common limit case of both general relativity and quantum mechanics (as described by figure 1). Hence, from this standpoint, what we used to call (before the 20th century) "classical" physics is an even weaker concept than what we would like to use. Even "classical" (using the modern interpretation) gauge interactions would therefore require some of the components of quantum mechanics. It seems like quantum mechanics and (apparently) classical general relativity must have a higher common theory that incorporates both of them as two dual sides of an over-arching theory. In figure 1, the relation appears somewhat like the tangent and co-tangent aspects of some over-arching theory, while the truly classical  (pre-quantum)  interpretation of nature would be a very narrow common limit of both these theories, expressible as a flat Minkowski spacetime with perfectly determined outcomes for all observables. Such a construction would only vaguely (and very approximately) fit to an empirical testing in the Natural world. 
Moreover, it seems like the gauge-quantum relation has some physical implications that can in principle be tested. For example, we have no reason not to expect quantum phenomena in situations in which $\hbar\rightarrow 0$, if the inner space gauge curvature term $\frac{\nabla^{2}R}{R}$ is large. In this sense, a limit $\frac{\hbar^{2}}{2m}\cdot \frac{\nabla^{2}R}{R}\rightarrow const.$ would equally well describe quantum effects, even if for the scale of our system, $\hbar$ remains relatively small. This aspect plays a significant role in the quantum interpretation of gluon confinement in QCD, as I will show in a future article.

\section{conclusion}
This article speculates about the quantum nature of gauge interactions, even without an explicit quantisation prescription applied on some classical theory. In fact I speculate that the gauge theory way of introducing interactions retains at least one aspect of quantum mechanics and in particular of quantum information, namely the non-cartesianity of the pairing of substructures. Slightly different from the way one combines wavefunctions with phases to create a system that has a space of states far larger than the cartesian pairing of the spaces of states of the individual subsystems, in the case of gauge fields, we try to combine patches of our base manifold (say spacetime) in order to produce a meaningful way to compare distant objects. Basically, in order to be able to do that, we also need to expand our space of fields, by accepting that two fields will be compared only up to a gauge transformation, allowing for a plethora of local choices. In order for those local choices to make sense globally, a gauge field and interaction emerges. However, this wouldn't be possible if one would pair the patches in a strict cartesian manner. The fibre bundle approach is something that allows us to go beyond cartesian pairing in gauge theories. In a sense it is precisely the role of our complex phase in quantum mechanics. In this sense, interaction appears as a form of "causally condensed" entanglement. Combining the adjacent patches in a consistent manner leaves us with the structures on each patch, namely, as we would say in quantum mechanics, with the constituent subsystems, but on top of that, also with a connection and hence an interaction, which compensates for the global structure of our manifold which is not directly visible locally, or, as we would call that in quantum terms, "entanglement". 
We also noted that there would be no interaction in a strictly classical world. Therefore Maxwell's equations not only  included in a hidden form the origins of special relativity and its upgraded causal structure, but also included in a hidden form, the idea of non-separability, intrinsic to quantum mechanics, through its gauge structure and the propagating nature of interactions. Of course, neither Maxwell nor Einstein could have noticed this, in the absence of the empirical evidence that led to quantum mechanics decades after both Maxwell's equations and Einstein's special relativity were constructed. It is interesting how historical chronology can affect the clarity of certain observations about the nature of reality. 
This approach makes the idea of spacetime emergence and particularly the idea of emergence of spacetime from entanglement more precise. 

\section{Data Availability and conflict of interests}
There is no conflict of interest.
Data sharing not applicable to this article as no datasets were generated or analysed during the current study.

\end{document}